\documentclass[sigconf]{acmart}

\usepackage{enumitem}
\usepackage{kotex}
\usepackage[linesnumbered,ruled,noend]{algorithm2e} 
\usepackage[switch]{lineno}
\usepackage{tabularx}
\usepackage{tikz}
\usepackage{pgfplots}
\usepackage{soul}
\usepackage{ragged2e}
\usepackage{array}
\usepackage{relsize}
\usepackage{bm}
\usepackage{graphicx}
\usepackage{latexsym}
\usepackage{subfig}
\usepackage{bm}

\pgfplotsset{compat=1.5.1}
\newenvironment{customlegend}[1][]{%
  \begingroup
  \csname pgfplots@init@cleared@structures\endcsname
  \pgfplotsset{#1}%
}{%
  \csname pgfplots@createlegend\endcsname
  \endgroup
}%
\def\addlegendimage{\csname pgfplots@addlegendimage\endcsname}
\usetikzlibrary{patterns}
\usetikzlibrary{pgfplots.groupplots}
\usetikzlibrary{
  pgfplots.colorbrewer,
}

\AtBeginDocument{%
  \providecommand\BibTeX{{%
    \normalfont B\kern-0.5em{\scshape i\kern-0.25em b}\kern-0.8em\TeX}}}

\copyrightyear{2022}
\acmYear{2022}
\setcopyright{acmcopyright}\acmConference[WSDM '22]{Proceedings of the Fifteenth
ACM International Conference on Web Search and Data Mining}{February 21--25,
2022}{Tempe, AZ, USA}
\acmBooktitle{Proceedings of the Fifteenth ACM International Conference on Web
Search and Data Mining (WSDM '22), February 21--25, 2022, Tempe, AZ, USA}
\acmPrice{15.00}
\acmDOI{10.1145/3488560.3498501}
\acmISBN{978-1-4503-9132-0/22/02}

\begin{document}
\fancyhead{} 
\title{Linear, or Non-Linear, That is the Question!}

\author{Taeyong Kong}
\authornote{Two first authors have contributed equally to this work.}
\affiliation{
	\institution{Yonsei University}
	\city{Seoul}
  	\country{Korea}
}
\email{qbxlvnf11@yonsei.ac.kr}

\author{Taeri Kim}
\authornotemark[1]
\affiliation{
	\institution{Hanyang University}
	\city{Seoul}
  	\country{Korea}
}
\email{taerik@hanyang.ac.kr}

\author{Jinsung Jeon}
\affiliation{
	\institution{Yonsei University}
	\city{Seoul}
  	\country{Korea}
}
\email{jjsjjs0902@yonsei.ac.kr}

\author{Jeongwhan Choi}
\affiliation{
	\institution{Yonsei University}
	\city{Seoul}
  	\country{Korea}
}
\email{jeongwhan.choi@yonsei.ac.kr}

\author{Yeon-Chang Lee}
\affiliation{
	\institution{Hanyang University}
	\city{Seoul}
  	\country{Korea}
}
\email{lyc0324@hanyang.ac.kr}

\author{Noseong Park}
\authornote{Co-corresponding authors.}
\affiliation{
	\institution{Yonsei University}
	\city{Seoul}
  	\country{Korea}
}
\email{noseong@yonsei.ac.kr}

\author{Sang-Wook Kim}
\authornotemark[2]
\affiliation{
	\institution{Hanyang University}
	\city{Seoul}
  	\country{Korea}
}
\email{wook@hanyang.ac.kr}



\begin{abstract}
There were fierce debates on whether the non-linear embedding propagation of GCNs is appropriate to GCN-based recommender systems. It was recently found that the linear embedding propagation shows better accuracy than the non-linear embedding propagation. Since this phenomenon was discovered especially in recommender systems, it is required that we carefully analyze the linearity and non-linearity issue. In this work, therefore, we revisit the issues of i) which of the linear or non-linear propagation is better and ii) which factors of users/items decide the linearity/non-linearity of the embedding propagation. We propose a novel \textbf{H}ybrid \textbf{M}ethod of \textbf{L}inear and non-lin\textbf{E}ar collaborative fil\textbf{T}ering method (\ours, pronounced as Hamlet). In our design, there exist both linear and non-linear propagation steps, when processing each user or item node, and our gating module chooses one of them, which results in a hybrid model of the linear and non-linear GCN-based collaborative filtering (CF). The proposed model yields the best accuracy in three public benchmark datasets. Moreover, we classify users/items into the following three classes depending on our gating modules' selections: Full-Non-Linearity (FNL), Partial-Non-Linearity (PNL), and Full-Linearity (FL). We found that there exist strong correlations between nodes' centrality and their class membership, \ie~important user/item nodes exhibit more preferences towards the non-linearity during the propagation steps. To our knowledge, we are the first who design a hybrid method and report the correlation between the graph centrality and the linearity/non-linearity of nodes. All {\ours} codes and
datasets are available at: https://github.com/qbxlvnf11/HMLET.
\end{abstract}

%


%


\newcommand{\ie}{{\it i.e.,}}
\newcommand{\eg}{{\it e.g.,}}
\newcommand{\yc}[1]{\textcolor{blue}{#1}}
\newcommand{\js}[1]{\textcolor{green}{#1}}
\newcommand{\note}[1]{\textcolor{red}{#1}}
\newcommand{\go}{Gowalla}
\newcommand{\yelp}{Yelp2018}
\newcommand{\amazon}{Amazon-Book}
\newcommand{\light}{LightGCN}
\newcommand{\lrgccf}{LR-GCCF}
\newcommand{\ngcf}{NGCF}
\newcommand{\Etype}{\textsc{\textsf{HMLET}}(All)}
\newcommand{\Btype}{\textsc{\textsf{HMLET}}(Front)}
\newcommand{\Ftype}{\textsc{\textsf{HMLET}}(Middle)}
\newcommand{\Dtype}{\textsc{\textsf{HMLET}}(End)}

\newcommand{\ourf}{\textsc{\textsf{HMLET}}(Front)}
\newcommand{\ourm}{\textsc{\textsf{HMLET}}(Middle)}
\newcommand{\oure}{\textsc{\textsf{HMLET}}(End)}

\newcommand{\ours}{\textsc{\textsf{HMLET}}}
\newcommand{\geee}{\textsc{\textsf{GEE-E}}}
\maketitle

\vspace{-0.2cm}
\section{Introduction}\label{sec:intro}
Recommender systems, personalized information filtering (IF) technologies, can be applied to many services, ranging from E-commerce, advertising, and social media to many other online and offline service platforms~\cite{Ying18pinsage}. One of the most popular recommender systems, collaborative filtering (CF), provides personalized preferred items to users by learning user and item embeddings from their historical user-item interactions~\cite{Ebesu2018, He17NeuMF, Hu08WRMF, Koren09MF, Rendle09BPR, Wang2014, ChaeKKL18, LeeK018, ChaeKCK20, choi2021ltocf}.

One of the mainstream research directions in recommender systems is how to learn high-order connectivity of user-item interactions while filtering out noises. Recently, GCN-based CF methods became popular in recommender systems because they show strong points to capture such latent high-order connectivity. Since existing GCNs are originally designed for graph or node classification tasks on attributed graphs, however, two limitations had been raised out when it comes to GCN-based CF methods: training difficulty~\cite{Wu2019SGC,Chen20LRGCCF,He20LightGCN} and over-smoothing~\cite{chen2020measuring,Chen20LRGCCF,He20LightGCN}.
Over-smoothing degrades the recommendation accuracy by considering the connectivity information too much~\cite{Chen20LRGCCF}. To overcome these problems, a couple of linear GCNs (linear embedding propagation-based GCNs) were proposed~\cite{Chen20LRGCCF,He20LightGCN}. These methods effectively alleviate the aforementioned two limitations and show superior performance over non-linear GCNs (non-linear embedding propagation-based GCNs). Even though linear GCNs show the state-of-the-art performance in many benchmark CF datasets, it is questionable in our opinion whether they can properly handle users and items with various characteristics and whether linear GCNs are consistently superior to non-linear GCNs in all cases. In addition, we are curious about, if one outperforms the other, which factors of graphs decide it.

To this end, we propose a \textbf{H}ybrid \textbf{M}ethod of \textbf{L}inear and non-lin\textbf{E}ar collaborative fil\textbf{T}ering (\ours, pronounced as Hamlet), a GCN-based CF method. {\ours} has the following key design points: i) We adopt a gating concept to decide between the linear and non-linear propagation for each node in a layer. ii) We perform residual prediction, where the embeddings from all layers are aggregated and used collectively for final predictions. Therefore, we let our gating modules decide which of the linear or non-linear propagation is used for a certain node at a certain layer instead of relying on manually designed architectures. This gating mechanism's key point is how to generate appropriate one-hot vectors. For this purpose, we adopt the Gumbel-softmax~\cite{Gumbel54,Maddison14sampling}.

To our knowledge, we are the first who combines the linear and non-linear embedding propagation in a systematic way, \ie~via the gating in our paper. Our gating mechanism can be considered as a sort of \emph{neural architecture search} (NAS) for the GCN-based CF method. However, our proposed mechanism is  more sophisticated because it provides the switching function for each user/item and the overall GCN architecture can be varied from a node's perspective to another.

\begin{figure}[!t]
    \centering
    \subfloat[ {\Etype} ]{\label{fig.a}\includegraphics[width=1\columnwidth]{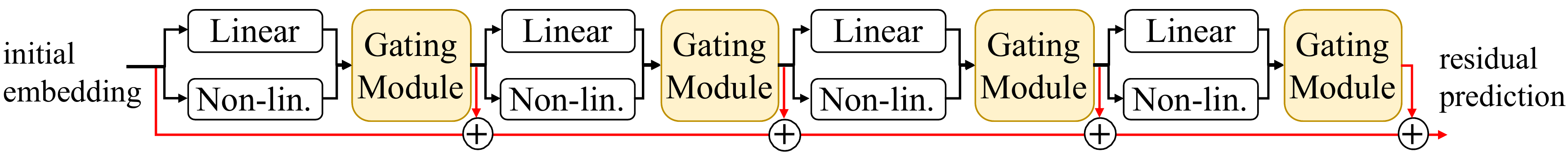}} \\
    \subfloat[ {\ourf} ]{\label{fig.b}\includegraphics[width=1\columnwidth]{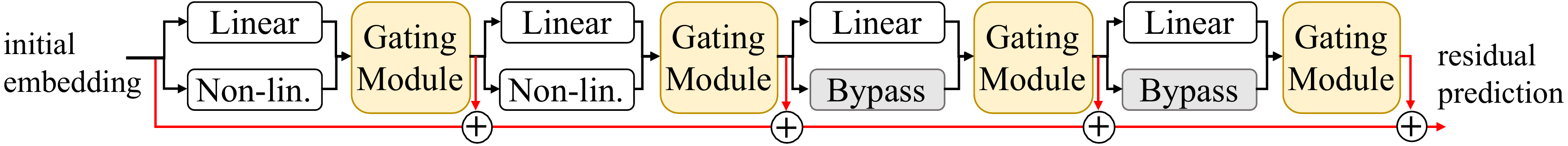}} \\
    \subfloat[{\ourm} ]{\label{fig.c}\includegraphics[width=1\columnwidth]{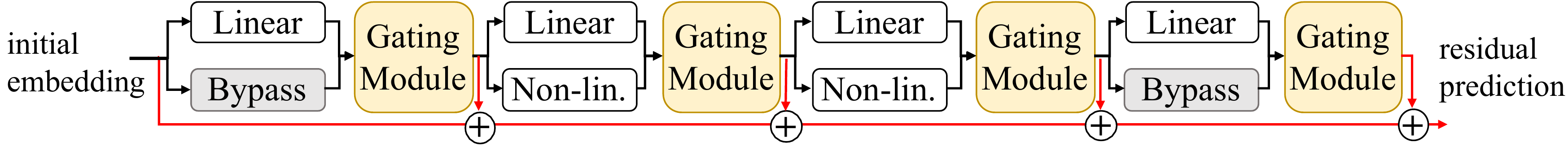}} \\
    \subfloat[{\oure} ]{\label{fig.d}\includegraphics[width=1\columnwidth]{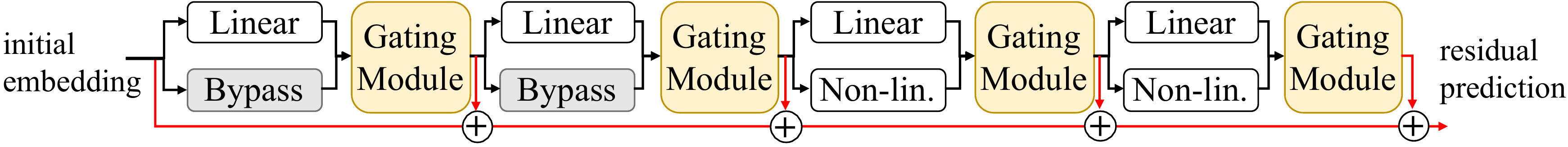}}
    \vspace{-0.3cm}
    \caption{Four variants of {\ours} in terms of the location of the non-linear propagation. {\Dtype} shows the best accuracy in our experiments. It was known that the problem of \emph{over-smoothing} happens with more than 2 non-linear propagation layers, and we use up to 2 non-linear layers.}\label{fig:hmlet}
    \vspace{-0.55cm}
\end{figure}

We conduct experiments with three benchmark CF datasets and compare our {\ours} with various state-of-the-art CF methods in terms of the normalized discounted cumulative gain (NDCG), recall, and precision. We also define several variations of {\ours} in terms of the locations of the non-linear propagation layers (see Fig.~\ref{fig:hmlet}). Among all of them, {\Dtype} shows the best performance in all datasets. Furthermore, we define three classes of nodes, \ie~users and items, depending on their preferences on the linear or non-linear propagation: Full-Non-Linearity (FNL), Partial-Non-Linearity (PNL), and Full-Linearity (FL). An FNL (resp. FL) node means that our gating module chooses the non-linear (resp. linear) propagation every time for the node and a PNL node has a mixed characteristic. At the end, we analyze the class-specific characteristics in terms of various graph centrality metrics and reveal that there exist strong correlations between the graph centrality, \ie~ the role of a node in a graph, and the linear/non-linear gating outcomes (see Table~\ref{tab:class1}). Our discovery shows that recommendation datasets are complicated because the linearity and non-linearity are mixed.

Contributions of our paper can be summarized as follows:
\begin{itemize}[leftmargin=*]
\item We propose {\ours}, which dynamically selects the best propagation method for each node in a layer.
\item We reveal that the role of a node in a graph is closely related to its linearity/non-linearity, \eg~ our gating module prefers the non-linear embedding propagation for the nodes with strong connections to other nodes.
\item Our experiments on three benchmark datasets show that {\ours} outperforms baselines in yielding better performance.
\end{itemize}

\begin{table}
  \caption{The characteristics of node classes in terms of various metrics. FNL (resp. FL) means a class of nodes for which our gating modules select only the non-linear (resp. linear) propagation in all layers. For PNL, our gating modules choose different propagation methods in different layers.}
  \vspace{-0.27cm}
  \small
  \label{tab:class1}
  \begin{tabular}{ccccc}
    \toprule
    Class & Degree & PageRank & Betweenness & Closeness \\
    \midrule
    FNL & High & High & High & High \\
    PNL & Moderate & Moderate & Moderate & Moderate \\
    FL & Low & Low & Low & Low \\
  \bottomrule
\end{tabular}
\vspace{-0.5cm}
\end{table}

\section{Related Works}\label{sec:related}
In this section, we review recommender systems and the Gumbel-softmax used in our proposed gating module.

\subsection{Recommender Systems}
Traditional recommender systems have focused on matrix factorization (MF) techniques~\cite{Koren09MF,HwangPKLL16}. Typical MF-based methods include BPR~\cite{Rendle09BPR} and WRMF~\cite{Hu08WRMF}. These MF-based methods simply learn relationships between users and items via dot products. Therefore, they have limitations in considering potentially complex relationships between users and items inherent in user-item interactions~\cite{He17NeuMF}. To overcome these limitations, deep learning-based recommender systems, \eg~Autoencoders~\cite{kingma2013auto,vincent2008extracting} and GCNs~\cite{Bruna2014,kipf2017semi,Wu2019SGC,Hamilton2017}, have been proposed to effectively learn more complicated relationships between users and items~\cite{Wang19NGCF,Ying18pinsage,zhang2020survey,Rianne17GCMC}.

Recently, recommender systems using GCNs~\cite{Wang19NGCF,Ying18pinsage,Rianne17GCMC} are gathering much attention. GCN-based methods can effectively learn the behavioral patterns between users and items by directly capturing the collaborative signals inherent in the user-item interactions~\cite{Wang19NGCF}. Typical GCN-based methods include GC-MC~\cite{Rianne17GCMC}, PinSage~\cite{Ying18pinsage}, and NGCF~\cite{Wang19NGCF}. In general, GCN-based methods model a set of user-item interactions as a user-item bipartite graph and then perform the following three steps:

{\bfseries (Step 1) Initialization Step:} They randomly set the initial $D$-dimensional embedding $\bm{e}^0$ of all user $u$ and item $v$, \ie\ $\bm  e_{u}^{0}, \bm e_{v}^{0} \in \mathbb {R}^{D}$. 


{\bfseries (Step 2) Propagation Step:} First of all, this propagation step is iterated $K$ times, \ie~ $K$ layers of embedding propagation. Given the $K$ layers, the embedding of a user node $u$ (resp. an item node $v$) in $i$-th layer is updated based on the embeddings of $u$'s (resp. $v$'s) neighbors $N_u$ (resp. $N_v$) in ($i-1$)-th layer as follows:

\vspace{-0.3cm}
\begin{equation}\label{eq:prop}
    \begin{split}
    \bm  e_{u}^{i} = \sigma ( \Sigma_{v\in N_u} \bm e_v^{i-1} \bm{W}_{i} ), \;\;\;     \bm  e_{v}^{i} = \sigma ( \Sigma_{u\in N_v} \bm e_u^{i-1} \bm{W}_{i} ),
    \end{split}
\end{equation}
where $\sigma$ denotes a non-linear activation function, \eg~ReLU, and $\bm W_{i} \in \mathbb {R}^{D \times D} $ is a trainable transformation matrix. There exist some other variations: i) including the self-embeddings, \ie~$N_u = N_u \cup \{u\}$ and $N_v = N_v \cup \{v\}$, ii) removing the transformation matrix, and iii) removing the non-linear activation function, which is in particular called as \emph{linear propagation}~\cite{Chen20LRGCCF, He20LightGCN}.

{\bfseries (Step 3) Prediction Step:} The preference of user $u$ to item $v$ is typically predicted using the dot product between the user $u$'s and item $v$'s embeddings in the last layer $K$, \ie~$\hat{r}_{u,v} = \bm e_{u}^{K} \odot \bm e_v^K$.

However, these GCN-based methods have two limitations: i) training difficulty ~\cite{Wu2019SGC,Chen20LRGCCF,He20LightGCN} and ii) over-smoothing, \ie~too similar embeddings of nodes~\cite{chen2020measuring,Chen20LRGCCF,He20LightGCN}. First, the training difficulty is caused by their use of a non-linear activation function in the propagation step~\cite{Wu2019SGC,Chen20LRGCCF,He20LightGCN}. Specifically, the non-linear activation function complicates the propagation step, and even worse, this operation is repeatedly performed whenever a new layer is created. Thus, they suffer from the training difficulty of the non-linear activation functions for large-scale user-item bipartite graphs ~\cite{Wu2019SGC,Chen20LRGCCF}. 

Next, the over-smoothing is caused as they use only the embeddings updated through the last layer in the prediction layer~\cite{Chen20LRGCCF}. Specifically, as the number of layers increases, the embedding of a node will be influenced more from its neighbors' embeddings. As a result, the embedding of a node in the last layer becomes similar to the embeddings of many directly/indirectly connected nodes~\cite{Chen20LRGCCF,chen2020measuring}. This phenomenon prevents most of the existing GCN-based methods from effectively utilizing the information of high-order neighborhood. Empirically, this is also shown by the fact that most of non-linear GCN-based methods show better performance when using only a few layers instead of deep networks.

Recently, LR-GCCF~\cite{Chen20LRGCCF} and LightGCN~\cite{He20LightGCN}, which are GCN-based recommender systems to alleviate the problems, have been proposed. First, to alleviate the former problem, they perform a linear embedding propagation without using a non-linear activation function in the propagation step. In order to mitigate the latter problem, they utilize the embeddings from all layers for prediction. After that, they perform \emph{residual prediction}~\cite{Chen20LRGCCF, He20LightGCN}, which predict each user's preference to each item with the multiple embeddings from the multiple layers. In~\cite{Chen20LRGCCF,He20LightGCN}, the authors demonstrated that a GCN architecture with the linear embedding propagation and the residual prediction can significantly improve the recommendation accuracy by successfully addressing the two problems.

\begin{table}
\footnotesize
\setlength{\tabcolsep}{2pt}
\center
  \caption{GCN-based recommender systems. In each layer, the gating module in {\ours} chooses either of the linear or the non-linear propagation for each node.}
  \vspace{-0.3cm}
  \label{tab:GCNs}
  \begin{tabular}{ccccccc}
  \toprule
  & GC-MC & PinSage & NGCF & LR-GCCF & LightGCN & \ours
\\ \midrule
Non-Linear Propagation & O & O & O & X & X & O \\
Linear Propagation & X & X & X & O & O & O \\
Residual Prediction & X & X & O & O & O & O\\
\bottomrule
\end{tabular}
\vspace{-0.8cm}
\end{table}

In summary, GCN-based recommender systems can be characterized by, as shown in Table~\ref{tab:GCNs}, the propagation and prediction types. We note that all existing methods consider only one of the linear or non-linear propagation, \ie~ they assume only one type of user-item interactions. However, we conjecture that user-item interactions are neither only linear nor only non-linear, for which we will conduct in-depth analyses in Section~\ref{sec:analysis}. In this paper, therefore, we propose a \textbf{H}ybrid \textbf{M}ethod of \textbf{L}inear and non-lin\textbf{E}ar collaborative fil\textbf{T}ering method (\ours), which considers both the two disparate propagation steps and selects an appropriate embedding propagation for each node in a layer.

\subsection{Gumbel-softmax}\label{subsec:gumbel}
The Gumbel-max trick~\cite{Gumbel54,Maddison14sampling} provides a way to sample a one-hot vector from a categorical distribution with class probabilities $\bm{\pi}$:
\begin{equation}
    \label{eqn:Gumbel-max}
   \bm{z}=one\_hot(\arg \max\limits_{\substack{i}}[g_i+\log\pi_i]),
\end{equation}
where $g_1...g_k$ are drawn from the unit Gumbel distribution. The $\arg \max$ operator does not allow the gradient flow via the Gumbel-max, because it gives zero gradients irrespective of how $\bm{\pi}$ was created. To this end, the Gumbel-softmax~\cite{jang16categorical} generates $\bm{y}$ that approximates $\bm{z}$ via the reparameterization trick defined as follows:
\begin{equation}
    \label{eqn:Gumbel-softmax}
    y_i=\frac{\exp \left(\left(\log \left(\pi _i\right)+g_i\right)/\tau \right)}{\sum _{j=1}^k\exp \left(\left(\log \left(\pi _j\right)+g_j\right)/\tau \right)}\;,
\end{equation}
where $y_i$ is $i$-th component of the vector $\bm{y}$, and $\tau$ is a temperature that determines how closely the function approximates $\bm{\pi}$. However, the Gumbel-softmax is challenging to use if it needs to sample discrete values because, when the temperature is high, its output is not categorical. To solve this problem, the straight-through Gumbel-softmax (STGS)~\cite{jang16categorical} can be used. STGS always generates discrete values for its forward pass, \ie~ $\bm{y}$ is a one-hot vector, while letting the gradients flow through $\bm{y}$ for its backward pass, even when the temperature is high. This makes neural networks with the Gumbel-softmax trainable.

This Gumbel-softmax has been widely used to learn optimal categorical distributions. One such example is network architecture search (NAS)~\cite{wu2019fbnet,he2020milenas,li2020neural,hu2020dsnas}. In NAS, we let an algorithm find optimal operators (among many pre-determined candidates prepared by users) and their connections. All these processes can be modeled by generating optimal one-hot (or multi-hot) vectors via the Gumbel-softmax~\cite{jang16categorical}. Another example is multi-generator-based generative adversarial networks (GANs)~\cite{goodfellow2014generative}. Park et al. showed that data is typically multi-modal, and it is necessary to separate modes and assign a generator to each mode of data, \eg~ one generator for long-hair females, another generator for short-hair males, and so on for a GAN generating facial images~\cite{Park2018}. In our case, we try to separate the two modes, \ie~ the linear and non-linear characteristics of nodes.

\section{Proposed Approach}\label{sec:method}

\begin{figure*}
  \centering
  \includegraphics[width=0.99\textwidth]{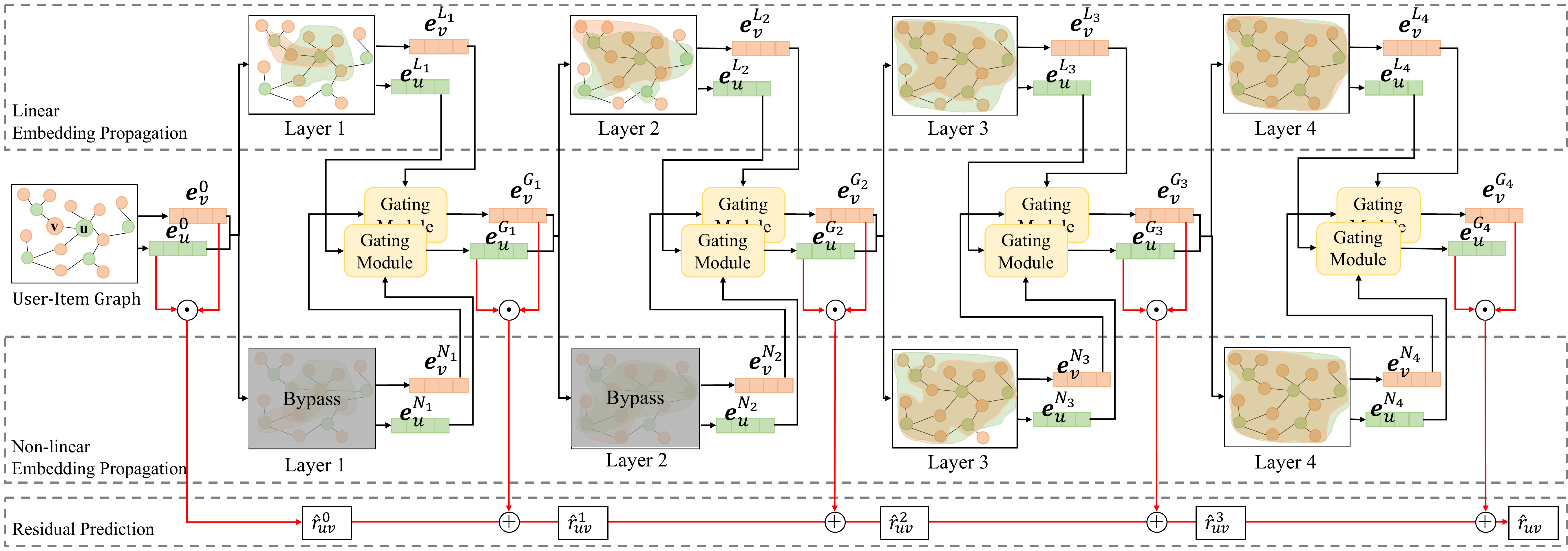}
  \vspace{-0.25cm}
  \caption{The detailed workflow of \oure. One can consider our gating module as a relay switch between the linear and non-linear propagation. While calculating an embedding for a user or an item in the third and fourth layer, therefore, our gating module learns the optimal selection between them for each node. For instance, it can select a sequence of linear $\rightarrow$ linear $\rightarrow$ linear $\rightarrow$ non-linear for some nodes while it can select a totally different sequence for other nodes.}\label{overview}
  \vspace{-0.3cm}
\end{figure*}
We first formulate our problem of top-$N$ recommendation as follows: Let $u \in U$ and $v \in I$ denote a user and an item, respectively, where $U$ and $I$ denote the sets of all users and all items, respectively; $N_u$ denotes a set of items rated by user $u$. For each user $u$, the goal is to recommend the top-\textit{$N$} items that are most likely to be preferred by $u$ among her unrated items, \ie\ $I \setminus N_u$.

In this section, among several variations of \ours, we mainly describe \oure~for ease of writing because it shows the best accuracy --- other variations can be easily modified from \oure~and we omit their descriptions. Its key concept is to adopt the gating between the linear and non-linear propagation in a layer. In other words, we prepare both the linear and non-linear propagation steps in a layer and let our gating module with STGS decide which one to use for each node. Table~\ref{table:notations} summarizes a list of notations used in this paper.

Figure~\ref{overview} illustrates the overall workflow of \oure. After constructing the user-item interaction as a user-item bipartite graph, \ours\ initializes the user and item embeddings in the initialization step. After that, each embedding is propagated to its neighbors through $K$ propagation layers. The gating module in \ours\ selects either of the linear or the non-linear propagation in a layer for each node (Section~\ref{sec:model-prop}). To this end, we use the gating module with STGS. In order to predict each user's preference on each item, the dot product of the user embedding and the item embedding in each layer is aggregated and we use their sum for prediction (Section~\ref{subsec:prediction}).

\begin{table}[t]
    \caption{Notations used in this paper}
    \vspace{-0.3cm}
    \footnotesize
    \centering
    \renewcommand{\arraystretch}{1.2}
    \begin{tabular}{c|l}
        \toprule
        \textbf{Notation}                 & \textbf{Description}    \\
        \midrule
        $K$  & The number of total layers \\
        $\bm{e}_{u}^{L_{i}}$, $\bm{e}_{v}^{L_{i}}$     & $u$'s and $v$'s embeddings at $i$-th linear layer      \\
        $\bm{e}_{u}^{N_{i}}$, $\bm{e}_{v}^{N_{i}}$       & $u$'s and $v$'s embeddings at $i$-th non-linear layer  \\
        $\bm{e}_{u}^{G_{i}}$, $\bm{e}_{v}^{G_{i}}$        & $u$'s and $v$'s embeddings selected by the gating module at $i$-th layer\\
        $D$    & The size (dimension) of embedding    \\
        $|U|$, $|I|$    & The number of users and items     \\
        $\hat{r}_{uv}$    & The user $u$'s final preference on item $v$    \\
        $\mathcal N_u$, $\mathcal N_v$    & The set of items rated to user $u$ and the set of users who rated item $v$ \\
        \bottomrule
    \end{tabular}
    \label{table:notations}
    \vspace{-0.55cm}
\end{table}





\subsection{Propagation Layer}\label{sec:model-prop}
We omit the description of the initialization step due to its obviousness. \ours\ propagates the embedding $\bm{e}_u$ (resp. $\bm{e}_v$) of each user $u$ (resp. each item $v$) through the propagation layers. In this subsection, we describe the propagation process. We first formally define the linear and the non-linear propagation steps used in this paper. Then, we present our gating module.

\subsubsection{\bfseries{Propagation}}

Recently, the authors of {\light}~\cite{He20LightGCN} found that the feature transformation and the non-linear activation do not have a positive effect on the effectiveness of CF. So, \light\ removed the feature transformation and the non-linear activation from Eq.~\eqref{eq:prop}, and it shows better performance than existing non-linear GCNs for recommendation. In \ours, we adopt the linear layer definition of \light. Therefore, our linear embedding propagation is performed as follows:

\small
\begin{equation}\label{eq:linear}
    \begin{split}
    \bm{e}_{u}^{L_{i+1}} = \mathop{\mathlarger{\sum}}_{v\in  \mathcal N_u} \frac{1}{\sqrt{\vert \mathcal N_u \Vert \mathcal N_v \vert}} \bm{e}_{v}^{G_{i}}, \;\;\;     \bm{e}_{v}^{L_{i+1}} = \mathop{\mathlarger{\sum}}_{u\in \mathcal N_v} \frac{1}{\sqrt{\vert \mathcal N_u \Vert \mathcal N_v \vert}} \bm{e}_{u}^{G_{i}},
    \end{split}
\end{equation}
\normalsize
where $\bm{e}_{u}^{L_{i+1}}$ and $\bm{e}_{v}^{L_{i+1}}$ are the linear embeddings for user $u$ and item $v$. Since our gating module, which will be described shortly, selects between the linear and the non-linear embeddings, $\bm{e}_{u}^{G_{i}}$ and $\bm{e}_{v}^{G_{i}}$ means the embeddings selected by our gating module in the previous $i$-th layer. If $i=0$, $\bm{e}_{u}^{G_{i}} = \bm{e}_{u}^0$ and $\bm{e}_{v}^{G_{i}} = \bm{e}_{v}^0$, \ie~ initial embeddings. $\frac{1}{\sqrt{\vert \mathcal N_u \Vert \mathcal N_v \vert}}$ is a symmetric normalization term to restrict the scale of embeddings into a reasonable boundary.

For the non-linear embedding propagation, we design a variant of the linear embedding propagation by adding non-linear activation functions. Its propagation is preformed as follows:
\vspace{-0.03cm}
\begin{equation}\label{eq:non-linear}
    \begin{split}
    \bm{e}_{u}^{N_{i+1}} = 
    \left\{
        \begin{array}{ll}
            \bm{e}_{u}^{N_{i}}, & \text{if~} bypass \\
            \phi \Big( \mathop{\mathlarger{\sum}}_{v\in \mathcal N_u} \frac{1}{\sqrt{\vert \mathcal N_u \Vert \mathcal N_v\vert }}  \bm{e}_{v}^{G_{i}} \Big), & \text{if~}propagate,
        \end{array}
     \right.\\
     \bm{e}_{v}^{N_{i+1}} = 
     \left\{
        \begin{array}{ll}
            \bm{e}_{v}^{N_{i}}, & \text{if~} bypass \\
            \phi \Big( \mathop{\mathlarger{\sum}}_{u\in \mathcal N_v} \frac{1}{\sqrt{\vert \mathcal N_u \Vert \mathcal N_v\vert }}  \bm{e}_{u}^{G_{i}} \Big), & \text{if~}propagate,
        \end{array}
     \right.\\
    \end{split}
\end{equation}
\normalsize
where $\phi$ is a non-linear activation function, \eg~ELU, Leaky ReLU. For instance, as shown in Figure 2, \oure\ bypasses the non-linearity propagation on the first and second layers to address the over-smoothing problem and then propagates the non-linear embedding in the third and fourth layers.


\subsubsection{\bfseries{Gating Module}}
Now, we have the two types of the embeddings for each node, created by the linear and non-linear propagation in Eqs.~\eqref{eq:linear} and~\eqref{eq:non-linear}, respectively, in the previous $i\text{-th}$ layer. Therefore, we should select one of the linear and non-linear embeddings for the propagation in the next $(i+1)\text{-th}$ layer. 

Toward this end, we add a gating module, which dynamically selects either of the linear or non-linear embedding after understanding the inherent characteristics of nodes. A separate gating module should be added whenever the linear and the non-linear propagation co-exist in a layer. The intuition behind this technique is that i) the embeddings of nodes may exhibit both the linearity and the non-linearity in their characteristics and ii) the linearity and the non-linearity of nodes may vary from one layer to another.



The process of the gating module with STGS is shown in Algorithm ~\ref{alg:gating}. For simplicity but without loss of generality, we use the symbol $\bm{e}^{L}$ and $\bm{e}^{N}$ to denote the linear and non-linear embeddings, respectively, after omitting other subscripts and superscripts. We support three gating types: i) choose the linear embedding (bypassing the non-linear propagation), ii) choose the non-linear embedding (bypassing the linear propagation), and iii) let the gating module choose one of them. The variable $\xi$ notates the gating type. If $\xi$ is the first or second type, a designated embedding type is selected. If $\xi$ is the third type, the input embeddings, \ie\ the linear and non-linear embedding, are concatenated and then passed to an MLP (multi-layer perceptron) (Lines 7 and 8 in Algorithm~\ref{alg:gating}). The result of the MLP is a logit vector $\bm{l}$, an input for STGS (Line 9).
The logit vector $\bm{l}$ corresponds to $\log \bm{\pi} $ explained in Section~\ref{subsec:gumbel}. $\bm{g}$ represents a linear or non-linear selection by the gating module, \ie~ $\bm{g}$ is a two-dimensional one-hot vector. Therefore, $\bm{e}^{G}$ is the same as either of $\bm{e}^{L}$ or $\bm{e}^{N}$ (Line 10).

\begin{algorithm}[t]
  \small
  \DontPrintSemicolon
  \SetAlgoLined
  \KwIn{Linear embedding $\bm{e}^{L}$, Non-linear embedding $\bm{e}^{N}$, Temperature $\tau$, Gating type $\xi$}
  \SetKwFunction{FMain}{Gating\_Module}
  \SetKwProg{Fn}{Function}{:}{end}
  \Fn{\FMain{$\bm{e}^{L}, \bm{e}^{N}, \tau$, $\xi$}}{
        \uIf{$\xi$ = linear}{
                $\bm{e}^{G} \gets \bm{e}^{L}$\;
            }
         \uElseIf{$\xi$ = non-linear}{
                $\bm{e}^{G} \gets \bm{e}^{N}$\;
            }
        \Else{
            $\bm{e}_{\text{concat}} \gets \bm{e}^{L} ||  \bm{e}^{N} $\;
            $\bm{l} \gets \text{MLP}(\bm{e}_{\text{concat}})$\;
            $\bm{g} \gets \text{STGS}(\bm{l}, \tau)$\;
            $\bm{e}^{G} \gets \bm{g} \cdot  [\bm{e}^{L},  \bm{e}^{N}]$\;
            }
        \KwRet {$\bm{e}^{G}$} \;
  }
  \caption{Gating Module}\label{alg:gating}
\end{algorithm}
\setlength{\textfloatsep}{5pt}%
\vspace{-0.3cm}
\subsubsection{\bfseries{Variants of HMLET}}
As shown in Figure~\ref{fig:hmlet} and Table~\ref{tbl:type}, there can be four variants of \ours, denoted as \Etype, \ourf, \ourm, and \oure,  depending on the locations of the non-linear layers.
Each method except \Etype\ uses up to 2 non-linear layers since it is known that more than 2 non-linear layers cause the problem of over-smoothing~\cite{Chen20LRGCCF}. Moreover, we test with various options of where to put them.
First, \ourf\ focuses on the fact that GCNs are highly influenced by close neighborhood, \ie\ in the first and second layers~\cite{tang15line}. 
Therefore, \ourf\ adopts the gating module in the front and uses only the linear propagation layers afterwards.
Second, \ourm\ only uses the linear propagation in the front and last and then adopts the gating module in the second and third layers.
Last, as the gating module is located in the third and fourth layers, \oure\ focuses on gating in the third and fourth layers --- our experiments and analyses show that \oure\ is the best among the four variations of the proposed method. We select $\bm{e}^{L_{3}}$ or $\bm{e}^{N_{3}}$ at the third layer and $\bm{e}^{L_{4}}$ or $\bm{e}^{N_{4}}$ at the fourth layer via the gating modules, respectively. If the linear embeddings are selected for a node in all layers, it is the same as using a linear GCN with $K=4$ for processing the node. If the non-linear embedding is selected for other node in all layers, it reduces to a non-linear GCN with $K=2$. Likewise, \oure~can be considered as a node-wise dynamic GCN (between the linear and non-linear propagation) with varying $K \in \{2,4\}$.

\begin{table}[t]
\footnotesize
\setlength{\tabcolsep}{1pt}
  \caption{Variants of \ours\ in terms of their setting for the non-linear propagation in Eq.~\eqref{eq:non-linear} and the gating type \boldsymbol{$\xi$}}\label{tbl:type}
  \vspace{-0.3cm}
  \label{tab:variants}
  \begin{tabular}{ccccc}
    \toprule
    Layer & 1  & 2  & 3 & 4 \\
    \midrule
    \Etype & \it{propagate}/\it{gating}  & \it{propagate}/\it{gating}  & \it{propagate}/\it{gating}  & \it{propagate}/\it{gating} \\
    \ourf & \it{propagate}/\it{gating}   & \it{propagate}/\it{gating}   & \it{bypass}/\it{linear}  & \it{bypass}/\it{linear} \\
    \ourm &  \it{bypass/linear} & \it{propagate/gating}  & \it{propagate/gating}  & \it{bypass}/\it{linear} \\
    \oure &  \it{bypass}/\it{linear} & \it{bypass}/\it{linear}  & \it{propagate}/\it{gating}   & \it{propagate}/\it{gating}  \\
  \bottomrule
\end{tabular}
\vspace{-0.4cm}
\end{table}



\SetInd{0.5em}{0.5em}
\begin{algorithm}[t]
\small
\SetAlgoLined
\DontPrintSemicolon
\KwIn{The number of total layers $K$, A bipartite graph $G$}
  \SetKwFunction{FMain}{\ours}
  \SetKwProg{Fn}{Function}{:}{end}
  \Fn{\FMain{$K, G$}}{
    Initialize $\bm{e}_u^{0}, \bm{e}_v^{0},~\text{for} ~\forall u, ~\forall v $\;
    $iter \leftarrow 0$ \;
    \While{the BPR loss is not converged}{
        $\tau \leftarrow 1.0 \exp(-0.001 \times iter)$\;\label{alg:tau}
        $\hat{r}_{u,v} \mathrel{{=}} \bm{e}_u^{0}  \odot\  \bm{e}_v^{0}, ~\text{for} ~\forall u, ~\forall v $ \;\label{alg:res1}
        \label{alg:anneal}
        \For{$i \leftarrow 1 \; \text{to} \; K$}{ 
            $\bm{e}_u^{L_{i}} , \bm{e}_v^{L_{i}} = \text{Eq}.~\eqref{eq:linear}, ~\text{for} ~\forall u, ~\forall v$ \;
            $\bm{e}_u^{N_{i}} , \bm{e}_v^{N_{i}} =  \text{Eq}.~\eqref{eq:non-linear}, ~\text{for} ~\forall u, ~\forall v$ \;
            $\bm{e}_u^{G_{i}} = \FuncSty{Gating\_Module}(\bm{e}_u^{L_{i}},\bm{e}_u^{N_{i}},\tau, \xi_{i}), ~\text{for} ~\forall u$\;
            $\bm{e}_v^{G_{i}} = \FuncSty{Gating\_Module}(\bm{e}_v^{L_{i}},\bm{e}_v^{N_{i}},\tau, \xi_{i}), ~\text{for} ~\forall v$\;
            $\hat{r}_{u,v} \mathrel{{+}{=}} \bm{e}_u^{G_{i}} \odot \bm{e}_v^{G_{i}}, ~\text{for} ~\forall u, ~\forall v$  \;\label{alg:res3}
        }
    Update $\bm{e}_u^{0}, \bm{e}_v^{0}$ with the BPR Loss for $\forall u, \forall v$\;\label{alg:update}
    Train the parameters of the gating modules with the BPR Loss\;\label{alg:update2}
    $iter \mathrel{{+}{=}} 1$ \;
    }
    \Return $\hat{r}_{u,v}$, for $\forall u, \forall v$\;
    }
\caption{\ours}
\label{algo:type_d} 
\end{algorithm}
\setlength{\textfloatsep}{5pt}%

\subsection{Prediction Layer}\label{subsec:prediction}
After propagating through all $K$ layers, we predict a user $u$'s preference for an item $v$. To this end, we create a dot product value of $\bm{e}^{G_{i}}_u$ and $\bm{e}^{G_{i}}_v$ in each layer and use the following residual prediction:

\small
\begin{equation}\begin{gathered}\label{eq:aaa}
    \hat{r}^{i}_{uv} = \bm e_{u}^{G_{i}} \odot \bm e_{v}^{G_{i}}, \;\;\;     \hat{r}_{uv} = \;\beta \mathop{\mathlarger{\sum}}^K_{i=0} \hat{r}^{i}_{uv}. 
\end{gathered}\end{equation}
\normalsize
In some layers, a gating module can be missing. In such a case, there is only one type of embeddings, but we also use $\bm{e}_{u}^{G_{i}}$/$\bm{e}_{v}^{G_{i}}$ to denote these embeddings for ease of writing.

In most previous GCN-based recommender system research, only the embedding of the last layer was used to predict, but in \ours, the above residual prediction $\hat r_{uv}$ with $\beta$ is used. Similar to LightGCN, $\beta$ is set to $1/(K+1)$. This residual prediction can produce good performance by using not only the embedding in the last layer but also the embeddings in previous layers.

\subsection{Training Method}\label{subsec:opt}

For training \ours, we employ the Bayesian Personalized Ranking (BPR) loss~\cite{Rendle09BPR}, denoted $\bm L$, which is frequently used in many CF methods. The BPR loss is written as follows:

\vspace{-0.4cm}
\small
\begin{equation}
    \begin{split}
    \bm L = - \mathop{\mathlarger{\sum}}^{|U|}_{u=1} \mathop{\mathlarger{\sum}}_{i\in N_u} \mathop{\mathlarger{\sum}}_{j\notin N_u}  \ln (\sigma(\hat{r}_{ui} - \hat{r}_{uj})) + \lambda\lVert \bm{\Theta} \rVert ^2,
    \end{split}
\end{equation}
\normalsize
where $\sigma$ is the sigmoid function. $\bf{\Theta}$ is the initial embeddings and the parameters of the gating modules, and $\lambda$ controls the $L_2$ regularization strength. We use each observed user-item interaction as a positive instance and employ the strategy used in~\cite{He20LightGCN} for sampling a negative instance.

We employ STGS for a smooth optimization of the gating module. We can train the network with annealing the temperature $\tau$, and we use the temperature decay for each epoch (Line~\ref{alg:tau} in Algorithm~\ref{algo:type_d}). In order to calculate $\hat{r}_{uv}$ as in Eq.~\eqref{eq:aaa}, we accumulate the dot product results (Lines~\ref{alg:res1} and~\ref{alg:res3}). Then, we train the initial embeddings (Line~\ref{alg:update}) and the parameters of the gating modules (Line~\ref{alg:update2}).


\section{Experiments}\label{sec:experiments}

In this section, we evaluate our proposed approach via comprehensive experiments. We design our experiments, aiming at answering the following key research questions (RQs):

{\begin{itemize}[leftmargin=*]
\item {\bfseries RQ1:} Which variation of {\ours} is the most effective in terms of recommendation accuracy?
\item {\bfseries RQ2:} Does gating between the linear and non-linear propagation provide more accurate recommendations than baseline methods?
\item {\bfseries RQ3:} 
What are the characteristics of the nodes that use i) only the linear propagation, ii) only the non-linear propagation, or iii) different propagation steps in different layers?
\end{itemize}}

\subsection{Experimental Environments}
\subsubsection{\bfseries{Datasets}}
For evaluation, we used the following three real-word datasets: {\go}, {\yelp}, and {\amazon} from various domains. They are all publicly available. Table~\ref{tab:datasets} shows the detailed statistics of the three datasets.

\begin{itemize}[leftmargin=*]
\item \textbf{\go} is a location-based social networking website where users share their locations by checking-in~\cite{liang2016modeling}. This dataset contains user-website interactions.

\item  \textbf{\yelp} is a subset of small business and user data used in Yelp Dataset Challenge 2018. 
This dataset contains user-business interactions.

\item  \textbf{\amazon} contains purchase records of Amazon users~\cite{he2016ups}. This dataset contains user-item interactions. {\amazon} has the highest sparsity among these three public datasets. 
\end{itemize}

\begin{table}[t]
\small
  \caption{Statistics of public benchmark datasets}
  \vspace{-0.3cm}
  \label{tab:datasets}
  \begin{tabular}{ccccl}
    \toprule
    Dataset & \# User & \# Item & \# Interaction & Sparsity \\
    \midrule
    \go & 29,858 & 40,981 & 1,027,370 & 99.916\%\\
    \yelp & 31,668 & 38,048 & 1,561,406 & 99.870\%\\
    \amazon & 52,643 & 91,599 & 2,984,108 & 99.938\%\\
  \bottomrule
\end{tabular}
\vspace{-0.4cm}
\end{table}

Following~\cite{Wang19NGCF}, we filtered out those users and items with less than ten interactions in all datasets, \ie\ a 10-core setting. For testing, we then split a dataset into training (80\%), validation (10\%), and test (10\%) sets in the same way as in ~\cite{Wang19NGCF}.

\subsubsection{\bfseries{Baseline Methods}}
We compare {\ours} with the following five state-of-the-art methods to verify its effectiveness:
\begin{itemize}[leftmargin=*]
    \item  \textbf{BPR}~\cite{Rendle09BPR} is a matrix factorization (MF) trained by the Bayesian Personalized Ranking (BPR) loss.
    \item \textbf{WRMF}~\cite{Hu08WRMF} is an MF solved by the weight alternating least square (WALS) technique.
    \item \textbf{\ngcf}~\cite{Wang19NGCF} is a non-linear GCN-based recommender system performing residual prediction.
    \item  \textbf{\lrgccf}~\cite{Chen20LRGCCF} is a linear GCN-based recommender system which removes the non-linear activation function but still use the transformation matrix in Eq.~\eqref{eq:prop}. This method performs the residual prediction.
    \item  \textbf{\light}~\cite{He20LightGCN} is yet another linear GCN-based recommender system performing the residual prediction. This method differs from {\lrgccf} in that it does not use the transformation matrix.
\end{itemize}

For MF-based methods, we use the implementations in the popular open-source library, called NeuRec.\footnote{\url{https://github.com/wubinzzu/NeuRec.}} 
For GCN-based methods, we use the source codes provided by the authors~\cite{He20LightGCN,Chen20LRGCCF,Wang19NGCF}.
To evaluate accuracy, we use the top-20 recommendations and measure the accuracy in terms of the normalized discounted cumulative gain (NDCG), recall, and precision, which are all frequently used in recommendation research~\cite{Wang19NGCF,Chen20LRGCCF,He20LightGCN}.

\subsubsection{\bfseries{Hyper-parameter Settings}}
We choose the best hyper-parameter set via the grid search with the validation set. The best setting found in {\ours} is as follows: the number of linear layers is set to 4; the number of non-linear layers is set to 4 in {\Etype} and 2 for {\Btype}, {\Ftype}, and {\Dtype}; the optimizer is Adam; the learning rate is 0.001; the $L_2$ regularization coefficient $\lambda$ is 1E-4; the mini-batch size is 2,048; the dropout rate is 0.4. And, we use the temperature $\tau$ with an initialization to 0.7, a minimum temperature of 0.01, and a decay factor of 0.995. Also, for fair comparison, we set the embedding sizes for all methods to 512. In non-linear layers, we test two non-linear activation functions: Leaky-ReLU (negative slope = 0.01) and ELU ($\alpha$ = 1.0). For baseline models, we tuned their hyper-parameters via the grid search in the ranges suggested in their respective papers.

\subsection{Experimental Results}

\begin{figure}[t!]
\centering
\begin{tikzpicture}
\centering
\footnotesize
\begin{customlegend}[legend columns=4,legend style={align=left,draw=none,column sep=0.5ex}, legend cell align={left},
  legend entries={{\Etype}, {\Btype}, {\Ftype}, {\Dtype}}]
  \addlegendimage{area legend, draw=black, fill=black} 
  \addlegendimage{area legend, draw=black, fill=black!70} 
  \addlegendimage{area legend, draw=black, fill=black!40} 
  \addlegendimage{area legend, draw=black, fill=black!10} 
  \end{customlegend}
\end{tikzpicture}

\begin{tabular}{c}  
\hspace{-1cm}
\begin{tikzpicture}
\footnotesize
  \centering
  \begin{axis}[
        title={\textbf{\amazon}},
        ybar=6pt, 
        height=2.8cm, width=.15\textwidth,
        bar width=0.2cm,
        x=1cm,
        ymajorgrids=true, 
        tick align=inside,
        major grid style={line width=.2pt,draw=white},
        enlarge x limits=0.15,
        ymin=0.0250, ymax=0.0300,
        axis x line*=center,
        axis y line*=left,
        ylabel={NDCG@20},
        symbolic x coords={Types},
        xtick=data,
        nodes near coords align=vertical,
        clip=false,
        scaled y ticks = false,
        y tick label style={/pgf/number format/fixed,
         /pgf/number format/precision=3},
    ]
    \addplot [draw=black, fill=black] coordinates {
      (Types, 0.0280)};
    \addplot [draw=black, fill=black!70] coordinates {
      (Types, 0.0280)};
    \addplot [draw=black, fill=black!40] coordinates {
      (Types, 0.0280)};
    \addplot [draw=black, fill=black!10] coordinates {
      (Types, 0.0300)};
  \end{axis}
\end{tikzpicture}
\begin{tikzpicture}
  \centering
  \footnotesize
  \begin{axis}[
        title={\textbf{\yelp}},
        ybar=6pt, 
        height=2.8cm, width=.15\textwidth,
        bar width=0.2cm,
        x=1cm,
        ymajorgrids=true, 
        tick align=inside,
        major grid style={line width=.2pt,draw=white},
        enlarge x limits=0.15,
        ymin=0.0390, ymax=0.0440,
        axis x line*=center,
        axis y line*=left,
        symbolic x coords={Types},
        xtick=data,
        nodes near coords align=vertical,
        clip=false,
        scaled y ticks = false,
        y tick label style={/pgf/number format/fixed,
         /pgf/number format/precision=3}, 
    ]
    \addplot [draw=black, fill=black] coordinates {
      (Types, 0.0431)};
    \addplot [draw=black, fill=black!70] coordinates {
      (Types, 0.0430)};
    \addplot [draw=black, fill=black!40] coordinates {
      (Types, 0.0430)};
    \addplot [draw=black, fill=black!10] coordinates {
      (Types, 0.0434)};
  \end{axis}
\end{tikzpicture}
\begin{tikzpicture}
  \centering
  \footnotesize
  \begin{axis}[
        title={\textbf{\go}},
        ybar=6pt, 
        height=2.8cm, width=.15\textwidth,
        bar width=0.2cm,
        x=1cm,
        ymajorgrids=true, 
        tick align=inside,
        major grid style={line width=.2pt,draw=white},
        enlarge x limits=0.15,
        ymin=0.1190, ymax=0.1240,
        axis x line*=center,
        axis y line*=left,
        symbolic x coords={Types},
        xtick=data,
        nodes near coords align=vertical,
        clip=false,
       scaled y ticks = false,
        y tick label style={/pgf/number format/fixed,
         /pgf/number format/precision=3},
    ]
    \addplot [draw=black, fill=black] coordinates {
      (Types, 0.1221)};
    \addplot [draw=black, fill=black!70] coordinates {
      (Types, 0.1221)};
    \addplot [draw=black, fill=black!40] coordinates {
      (Types, 0.1219)};
    \addplot [draw=black, fill=black!10] coordinates {
      (Types, 0.1231)};
  \end{axis}
\end{tikzpicture}
\end{tabular}
\vspace{-0.4cm}
\caption{The comparison of NDCG@20 with all types of {\ours} in three public benchmarks.}
\label{fig:comparison-rq1}
\end{figure}
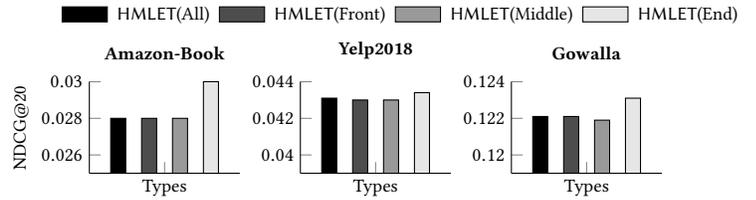

\subsubsection{\bfseries{Comparison among Model Variations (RQ1)}}
For answering RQ1, we first compare the accuracies of \Etype, \Btype, \Ftype, and \Dtype. Figures~\ref{fig:comparison-rq1} illustrates the results 
where X-axis represents the types of {\ours}, and Y-axis represents NDCG@20. 

\begin{table*}[t]
\caption{The comparison of overall performance with baseline models on three public benchmarks}
\vspace{-0.3cm}
\label{tab:comparison-rq1}
\resizebox{0.95\textwidth}{!} {%
\begin{tabular}{|c|c|c|c|c|c|c|c|c|c|}
\toprule
  & \multicolumn{3}{c|}{Amazon-Book}                                                                                                                                                                       & \multicolumn{3}{c|}{Yelp2018}                                                                                                                                                                         & \multicolumn{3}{c|}{Gowalla}                                                                                                                                                                          \\ \hline
Metrics   & NDCG@20                                                    & Recall@20                                                  & Precision@20                                                  & NDCG@20                                                       & Recall@20                                                    & Precision@20                                                & NDCG@20                                                     & Recall@20                                                    & Precision@20\\                  \hline\hline BPR      & 0.0181                                                              & 0.0307                                                              & 0.0065                                                              & 0.0289                                                              & 0.0476                                                              & 0.0062                                                              & 0.0847                                                              & 0.1341                                                              & 0.0203                                                              \\ \hline
WRMF     & 0.0242                                                            & 0.0405                                                            & 0.0085                                                            & 0.0402                                                              & 0.0645                                                              & 0.0145                                                              & 0.1007                                                              & 0.1538                                                              & 0.0243        \\ \hline
NGCF     & 0.0250                                                            & 0.0426                                                            & 0.0086                                                              & 0.0371                                                            & 0.0601                                                            & 0.0134                                                            & 0.1071                                                            & 0.1661                                                            & 0.0253                                                            \\ \hline
LR-GCCF  &  0.0213                                                           & 0.0361                                                            & 0.0077                                                            & 0.0351                                                            & 0.0581                                                            & 0.0129                                                            & 0.0989                                                            & 0.1545                                                            & 0.0240                                                           \\ \hline
{\light} & \textit{0.0283}                                                              & \textit{0.0484}                                                             & \textit{0.0099}                                                              & \textit{0.0420}                                                              & \textit{0.0678}                                                              & \textit{0.0152}                                                              & \textit{0.1212}                                                              & \textit{0.1870}                                                              & \textit{0.0288}                                                              \\ \hline\hline

\textbf{{\oure}}    & \textbf{\begin{tabular}[c]{@{}c@{}}0.0300\end{tabular}} & \textbf{\begin{tabular}[c]{@{}c@{}}0.0510\end{tabular}} & \textbf{\begin{tabular}[c]{@{}c@{}}0.0103\end{tabular}} & \textbf{\begin{tabular}[c]{@{}c@{}}0.0434\end{tabular}} & \textbf{\begin{tabular}[c]{@{}c@{}}0.0696\end{tabular}} & \textbf{\begin{tabular}[c]{@{}c@{}}0.0155\end{tabular}} & \textbf{\begin{tabular}[c]{@{}c@{}}0.1231\end{tabular}} & \textbf{\begin{tabular}[c]{@{}c@{}}0.1908\end{tabular}} & \textbf{\begin{tabular}[c]{@{}c@{}}0.0293\end{tabular}} \\ \hline 
\%Improve      & 6.00\%                                                              & 5.37\%                                                              & 4.04\%                                                              & 3.33\%                                                              & 2.65\%                                                              & 1.97\%                                                              & 1.56\%                                                              & 2.03\%                                                             & 1.73\%                                                             \\ \hline
$p$-value     & 1.59E-37
                                                            & 1.71E-34
                                                            & 5.16E-36
                                                            &  5.57E-58
                                                             & 2.90E-39

                                                              & 1.61E-44
                                                              & 8.51E-25                                                              & 4.17E-131                                                              & 3.39E-27                      \\

\bottomrule\hline
\end{tabular}
}
\vspace{-0.2cm}
\end{table*}

\begin{table}[t]
\footnotesize
\setlength{\tabcolsep}{1pt}
  \caption{The selection ratio by gating modules in {\amazon}}
  \vspace{-0.3cm}
  \label{tab:gating_type}
  \begin{tabular}{|c|c|c|c|c|c|c|c|c|}
    \toprule
    &    \multicolumn{2}{c|}{Layer 1}&    \multicolumn{2}{c|}{Layer 2}&   
    \multicolumn{2}{c|}{Layer 3}&   
    \multicolumn{2}{c|}{Layer 4}\\   
    \hline
    Propagation & Linear & Non-Lin. & Linear & Non-Lin. & Linear & Non-Lin. & Linear & Non-Lin. \\
    \hline
    {\Etype} & 77.05\% & 22.95\% & 60.45\% & 39.55\% & 68.01\% & 31.99\% & 40.77\% & 59.23\% \\
    {\Btype} & 74.30\% & 25.70\% & 62.09\% & 37.91\% & 100\% & 0\% & 100\% & 0\%\\
    {\Ftype} & 100\% & 0\% & 63.03\% & 36.97\% & 72.41\% & 27.59\% & 100\% & 0\%\\
    {\Dtype} &100\% &0\%  &100\%  &0\%  &       5.19\% & 94.81\% & 49.55\% & 50.45\%\\
  \bottomrule
\end{tabular}
\end{table}

{\Dtype} is the best among all variations of {\ours}. The accuracies of all variations except {\Dtype} are similar. Specifically, the difference between {\Dtype} and other variations is around 7\%, 0.9\%, 1\% in {\amazon}, {\yelp}, and {\go}, respectively. However, the differences in accuracy among {\Etype}, {\Btype}, and {\Ftype} are as small as 0.2\%.

These results indicate that i) the effectiveness of the gating module greatly depends on the location where the gating module exists, and ii) the non-linear propagation is useful to capture distant neighborhood information
--- note that we added the gating modules at the last two layers in {\Dtype}. As shown in Table~\ref{tab:gating_type}, each variation has a quite different linear/non-linear embedding selection ratio. {\Dtype}, the best model, uses the non-linear propagation in the layers 3 and 4, and their selection ratios are significant, \eg~ 5.19\% of linear vs. 94.81\% of non-linear in the third layer. This observation also applies to the second best model, {\Etype}. Similar selection ratio patterns are observed in the other two datasets.

\vspace{-0.05cm}
\subsubsection{\bfseries{Comparison with Baselines (RQ2)}} \label{sec:comparison_competing_methods}
Table~\ref{tab:comparison-rq1} illustrates our main experimental results.
In them, the values in boldface indicate the best accuracy in each column, and the values in italic mean the best baseline accuracy. 
Also, `\%Improve' indicates the degree of accuracy improvements over the best baseline by \oure.
Lastly, we conduct $t$-tests with a 95\% confidence level to verify the statistical significance of the accuracy differences between {\oure} and the baselines.

We summarize the results shown in Table~\ref{tab:comparison-rq1}  as follows. 
First, among the five baseline methods, we observe that {\light} consistently shows the best accuracy in all datasets.
Second, {\oure} consistently provides the highest accuracy in all datasets and with all metrics. Specifically, {\oure} outperforms {\light} by 6.00\%, 3.33\%, and 1.56\% for the datasets in terms of NDCG@20, respectively. The $p$-values are below $0.05$, indicating that the differences are statistically significant. We highlight that {\oure} shows remarkable improvements in {\amazon} which is the largest dataset in this paper.


\vspace{-0.1cm}
\subsection{Linearity or Non-Linearity}\label{sec:analysis}
In this subsection, we define three different classes of nodes, depending on their preferences on the linear or non-linear propagation, and perform in-depth analyses on them. In order to analyze accurately, we use the embeddings learned by {\Dtype}, the best performing variation of \ours, for {\amazon}. Due to space limitations, we omit the results for the other datasets, which show similar patterns to those in {\amazon}. 

\subsubsection {\bfseries{Node Class and Graph Centrality}}\label{sec:node_class}
We first classify all nodes into one of the following three classes according to the embedding types selected by the gating modules:

\begin{itemize}[leftmargin=*]
    \item {\bfseries Full-Non-Linearity (FNL)} is a class of nodes in which all embeddings selected by the gating modules are non-linear embeddings.

\item {\bfseries Partial-Non-Linearity (PNL)} is a class of nodes in which the embeddings selected by the gating modules are mixed with linear embeddings and non-linear embeddings.

\item {\bfseries Full-Linearity (FL)} is a class of nodes in which all embeddings selected by the gating modules are linear embeddings.
\end{itemize}

We next introduce three graph centrality metrics to study the characteristics of the classes:

\begin{itemize}[leftmargin=*]

\item {\bfseries PageRank} 
~\cite{page1999pagerank} measures the relative importance of nodes in a graph. A node is considered as important, even though its connectivity with other nodes is not that strong, if connected to other important nodes.

\item {\bfseries Betweenness Centrality} ~\cite{brandes2001centralitybet} measures the centrality of a node as an intermediary in a graph. The more a node appears in multiple shortest paths, the higher the betweenness centrality of the node. 

\item {\bfseries Closeness Centrality}~\cite{freeman1978centralitycls}
measures the centrality of a node by considering general connections to other nodes in a graph. The less hops it takes for a node to reach all other nodes, the higher the closeness centrality.
\end{itemize}

\subsubsection {\bfseries{Characteristics of Node Classes (RQ3)}}\label{sec:node_class_ratio}
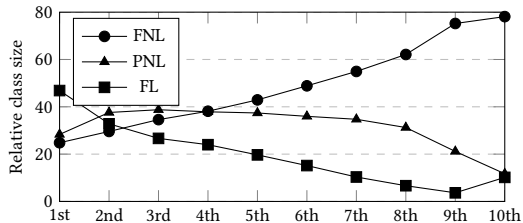
\begin{figure}
\footnotesize
\centering
\begin{tikzpicture}
\centering
\begin{axis}[
    height=4.1cm, width = 7.5cm,
    ylabel={Relative class size},
    xmin=0, xmax=90,
    ymin=0, ymax=80,
    xtick={0,10,20,30,40,50,60,70,80,90},
    xticklabels={1st, 2nd, 3rd, 4th, 5th, 6th, 7th, 8th, 9th, 10th},
    ytick={0,20,40,60,80},
    legend pos=north west,
    ymajorgrids=true,
    grid style=dashed,
    mark options={solid}
]
\addplot[
   mark=*
    ]
    coordinates {
    (90,78.11895919) (80,75.20940871)
		(70,62.10739659) (60,54.9429539) (50,48.86218309) (40,42.92219384)
		(30,38.13066353) (20,34.57854787) (10,29.58106313) (0,24.81291262)
    };
\addplot[
    mark=triangle*
    ]
    coordinates {
    (90,11.70514673) (80,21.1389107)
		(70,31.2421691) (60,34.72602399) (50,35.97356713) (40,37.39357856)
		(30,37.8853709) (20,38.7649476) (10,37.60506112) (0,28.32400165)
    };
\addplot[
    mark=square*
    ]
    coordinates {
    (90,10.17589408) (80,3.651680582)
		(70,6.65043431) (60,10.33102211) (50,15.16424978) (40,19.68422761)
		(30,23.98396556) (20,26.65650453) (10,32.81387575) (0,46.86308574)
    };
\legend{FNL,PNL,FL}
\end{axis}
\end{tikzpicture}
\vspace{-0.3cm}
\caption{The class ratio of nodes sorted by degree. $i$-th bin in X-axis means a range of [\boldsymbol{$(10*(i-1))$}-th percentile, \boldsymbol{$(10*i)$}-th percentile) in terms of degree. Nodes with a high degree are the most likely to be in FNL (10th), and nodes with a small degree are likely to be in FL (1st). We also find that nodes classified as PNL are more evenly distributed than other classes.}
\label{fig:node_degree_distribution}
\end{figure}

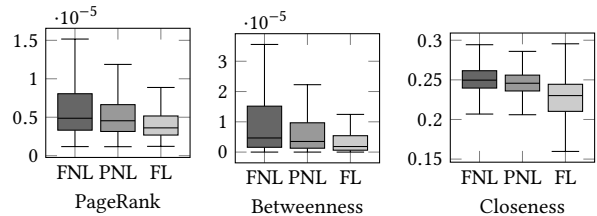
\begin{figure}[t!]
\small
\centering
\usepgfplotslibrary{statistics}
\begin{tikzpicture}
  \begin{axis}[
    height=3.3cm, width = 3.5cm,
    xlabel = {PageRank},
    boxplot/draw direction=y,
    xtick={1,2,3},
    xticklabels={FNL, PNL, FL},
    x tick label style={text width = 1cm, align=center}
    ]
    \addplot+[
    boxplot prepared={
      lower whisker=1.18196E-06,
      lower quartile=3.3164E-06,
      median=4.8673E-06,
      upper quartile=8.05604E-06,
      upper whisker=1.51649E-05,
    }, color = black, fill=black!60
    ] coordinates{};
    \addplot+[
    boxplot prepared={
      lower whisker=1.16982E-06,
      lower quartile=3.14872E-06,
      median=4.54145E-06,
      upper quartile=6.63666E-06,
      upper whisker=1.18678E-05
    }, color = black, fill=black!40
    ] coordinates{};
    \addplot+[
    boxplot prepared={
      lower whisker=1.21219E-06,
      lower quartile=2.68252E-06,
      median=3.61638E-06,
      upper quartile=5.16359E-06,
      upper whisker=8.87023E-06,
    }, color = black, fill=black!20
    ] coordinates{};
    \end{axis}
\end{tikzpicture}
\begin{tikzpicture}
  \begin{axis}[
    height=3.3cm, width = 3.5cm,
    xlabel = {Betweenness},
    boxplot/draw direction=y,
    xtick={1,2,3},
    xticklabels={FNL, PNL, FL},
    x tick label style={text width = 1cm, align=center}
    ]
    \addplot+[
    boxplot prepared={
      lower whisker=0,
      lower quartile=1.5827E-06,
      median=4.66162E-06,
      upper quartile=1.51682E-05,
      upper whisker=3.55464E-05,
    }, color = black, fill=black!60
    ] coordinates{};
    \addplot+[
    boxplot prepared={
      lower whisker=0,
      lower quartile=1.28016E-06,
      median=3.50505E-06,
      upper quartile=9.67357E-06,
      upper whisker=2.22579E-05
    }, color = black, fill=black!40
    ] coordinates{};
    \addplot+[
    boxplot prepared={
      lower whisker=0,
      lower quartile=6.21681E-07,
      median=1.83557E-06,
      upper quartile=5.4101E-06,
      upper whisker=1.24468E-05,
    }, color = black, fill=black!20
    ] coordinates{};
    \end{axis}
\end{tikzpicture}
\begin{tikzpicture}
  \begin{axis}[
    height=3.3cm, width = 3.5cm,
    xlabel ={Closeness},
    boxplot/draw direction=y,
    xtick={1,2,3},
    xticklabels={FNL, PNL, FL},
    x tick label style={text width = 1cm, align=center}
    ]
    \addplot+[
    boxplot prepared={
      lower whisker=0.206843691,
      lower quartile=0.239632048,
      median=0.24964174,
      upper quartile=0.261528364,
      upper whisker=0.294344759,
    }, color = black, fill=black!60
    ] coordinates{};
    \addplot+[
    boxplot prepared={
      lower whisker=0.205991771,
      lower quartile=0.235977482,
      median=0.245726143,
      upper quartile=0.255969292,
      upper whisker=0.285935459
    }, color = black, fill=black!40
    ] coordinates{};
    \addplot+[
    boxplot prepared={
      lower whisker=0.15967715,
      lower quartile=0.21024025,
      median=0.230146713,
      upper quartile=0.244418274,
      upper whisker=0.295508604,
    }, color = black, fill=black!20
    ] coordinates{};
    \end{axis}
\end{tikzpicture}
\vspace{-0.3cm}
\caption{The statistics of PageRank, betweenness centrality, and closeness centrality.
}
\vspace{-0.3cm}
\label{fig:box_plot}
\end{figure}

In this subsection, we analyze the characteristics of the nodes in each class in terms of the various centrality metrics.
Table~\ref{tab:class ratio} shows the relative class size in our three datasets. In {\amazon} and {\yelp}, most nodes were classified as FNL and PNL (about 47-48\% and 50-52\%, respectively), and a few nodes were classified as FL (about 1-2\%). However, in {\go}, the ratio of FL is about 12\%, which is relatively higher compared to the other two datasets.
Figure~\ref{fig:node_degree_distribution} shows the relative sizes of the three classes by degree, and Figure~\ref{fig:box_plot} shows the statistics of the centrality scores in each class. From them, it can be seen that the degree and centrality scores increase in order of FL, PNL, and FNL. Now, we deliver the meaning of the above results for each class.

\begin{table}
\small
  \caption{The relative class sizes in three datasets}
  \vspace{-0.3cm}
  \label{tab:class ratio}
  \begin{tabular}{cccc}
    \toprule
    Class & Amazon-Book & Yelp2018 & Gowalla \\
    \midrule
    FNL & 46.78\% & 47.95\% & 27.49\% \\
    PNL & 51.77\% & 49.59\% & 60.84\% \\
    FL & 1.45\% & 2.46\% & 11.67\% \\
  \bottomrule
\end{tabular}
\vspace{-0.1cm}
\end{table}

\begin{itemize}[leftmargin=*]
\item {\bfseries FNL Attributes:} 
 A node in FNL is either an active user or a popular item with more direct/indirect interaction information, \ie~a high degree and closeness centrality, and higher influence, \ie~a high PageRank and betweenness centrality, than nodes in other classes.
 So, they will receive a lot of information during the propagation step. Therefore, the sophisticated non-linear propagation is required to correctly extract useful information from much potentially noisy information.
 
 \item {\bfseries FL Attributes:}
 A node in FL is either a user or an item that does not have much direct/indirect interaction information, \ie~a low degree and closeness centrality, and little influence, \ie~a low PageRank and betweenness centrality, compared to nodes in other classes.
The information they receive during the propagation step may mostly consist of useful information related to themselves with little noise. Therefore, the simple linear propagation is required to take useful information as it is, rather than refining it.
 
 \item {\bfseries PNL Attributes:}
 A node in PNL, compared to nodes in other classes, is a user or an item with neither too large nor too small direct/indirect interaction information, \ie~a moderate degree and closeness centrality, and influence, \ie~a moderate PageRank and betweenness centrality. In other words, although they have many direct neighbors, there are few indirect neighbors connected to the direct neighbors, or even if there are few direct neighbors, their indirect neighbors can be many. Therefore, they need to perform one of the non-linear or linear operations depending on the information they receive from neighbors.
\end{itemize}
\vspace{2mm}

In order to double-check our interpretations, we show the statistics of the similarity of embeddings for each class. So, we calculate the cosine similarity between a node and its direct neighbors by using the embeddings learned by {\Dtype}. The results are shown in Table~\ref{tab:similarity}. From these results, we can confirm that the neighbors of a node in FNL consist of diversified nodes, \ie~a low mean and high variance. Also, FL nodes' neighbors mainly consist of similar nodes, \ie~a high mean and low variance. Lastly, PNL nodes' neighbors are in between the previous two cases, \ie~a moderate mean and variance between nodes in other classes.



\begin{table}[t]
\small
    \caption{The statistics of the cosine similarity between a node's embedding and its direct neighbors' embeddings in each node class}
    \vspace{-0.3cm}
    \label{tab:similarity}
    \begin{tabular}{cccc}
        \toprule
        Similarity & FNL & PNL & FL \\
        \midrule
        Mean & 0.6369 & 0.6449 & 0.7157 \\
        Variance & 0.0179 & 0.0162 & 0.0160 \\
        \bottomrule
    \end{tabular}
\end{table}

\section{Conclusions and Future Work}
In this paper, we presented a novel GCN-based CF method, named as {\ours}, that can select the linear or non-linear propagation step in a layer for each node. We further analyzed how the linear/non-linear selection mechanism works using various graph analytics techniques.
To this end, we first designed our linear and non-linear propagation steps, being inspired by various state-of-the-art linear and non-linear GCNs for CF. Then, we used STGS to learn the optimal selection between the linear and non-linear propagation steps. 
The intuition behind such design choice is that it is not optimal to put both the linear and the non-linear propagation in every layer. 
In this sense, we have defined several variations of \ours\ in terms of combining the linear and non-linear propagation steps. 

Through extensive experiments using three standard benchmark datasets, we demonstrated that \ours\ shows the best accuracy in all datasets. Furthermore, we presented in-depth analyses of how the linearity and non-linearity of nodes are decided in a graph. Toward this end, we classified nodes into three classes, \ie~Full-Non-Linearity, Partial-Non-Linearity, and Full-Linearity, depending on our gating module’s selections and studied correlations between nodes’ centrality scores and their class membership.

We conjecture that GCNs for CF should somehow consider both linear and non-linear operations. We do not say that our specific mechanism to combine the linear and the non-linear propagation steps is optimal. We hope that our discovery encourages much follow-up research work.
\label{sec:conclusions}

\section*{Acknowledgment}
The work of Sang-Wook Kim was supported by Samsung Research Funding \& Incubation Center of Samsung Electronics under Project Number SRFC-IT1901-03. The work of Noseong Park was supported by the Yonsei University Research Fund of 2021 and the Institute of Information \& Communications Technology Planning \& Evaluation (IITP) grant funded by the Korean government (MSIT) (No. 2020-0-01361, Artificial Intelligence Graduate School Program (Yonsei University)).


\bibliographystyle{ACM-Reference-Format}
\bibliography{bibliography}


\begin{thebibliography}{39}


\ifx \showCODEN    \undefined \def \showCODEN     #1{\unskip}     \fi
\ifx \showDOI      \undefined \def \showDOI       #1{#1}\fi
\ifx \showISBNx    \undefined \def \showISBNx     #1{\unskip}     \fi
\ifx \showISBNxiii \undefined \def \showISBNxiii  #1{\unskip}     \fi
\ifx \showISSN     \undefined \def \showISSN      #1{\unskip}     \fi
\ifx \showLCCN     \undefined \def \showLCCN      #1{\unskip}     \fi
\ifx \shownote     \undefined \def \shownote      #1{#1}          \fi
\ifx \showarticletitle \undefined \def \showarticletitle #1{#1}   \fi
\ifx \showURL      \undefined \def \showURL       {\relax}        \fi
\providecommand\bibfield[2]{#2}
\providecommand\bibinfo[2]{#2}
\providecommand\natexlab[1]{#1}
\providecommand\showeprint[2][]{arXiv:#2}

\bibitem[\protect\citeauthoryear{Brandes}{Brandes}{2001}]%
        {brandes2001centralitybet}
\bibfield{author}{\bibinfo{person}{Ulrik Brandes}.}
  \bibinfo{year}{2001}\natexlab{}.
\newblock \showarticletitle{A faster algorithm for betweenness centrality}.
\newblock \bibinfo{journal}{\emph{Journal of mathematical sociology}}
  \bibinfo{volume}{25}, \bibinfo{number}{2} (\bibinfo{year}{2001}),
  \bibinfo{pages}{163--177}.
\newblock


\bibitem[\protect\citeauthoryear{Bruna, Zaremba, Szlam, and LeCun}{Bruna
  et~al\mbox{.}}{2014}]%
        {Bruna2014}
\bibfield{author}{\bibinfo{person}{Joan Bruna}, \bibinfo{person}{Wojciech
  Zaremba}, \bibinfo{person}{Arthur Szlam}, {and} \bibinfo{person}{Yann
  LeCun}.} \bibinfo{year}{2014}\natexlab{}.
\newblock \showarticletitle{{Spectral networks and deep locally connected
  networks on graphs}}. In \bibinfo{booktitle}{\emph{Proc. of the Int'l Conf.
  on Learning Representations (ICLR)}}.
\newblock


\bibitem[\protect\citeauthoryear{Chae, Kang, Kim, and Lee}{Chae
  et~al\mbox{.}}{2018}]%
        {ChaeKKL18}
\bibfield{author}{\bibinfo{person}{Dong{-}Kyu Chae}, \bibinfo{person}{Jin{-}Soo
  Kang}, \bibinfo{person}{Sang{-}Wook Kim}, {and} \bibinfo{person}{Jung{-}Tae
  Lee}.} \bibinfo{year}{2018}\natexlab{}.
\newblock \showarticletitle{{CFGAN:} {A} Generic Collaborative Filtering
  Framework based on Generative Adversarial Networks}. In
  \bibinfo{booktitle}{\emph{Proc. of the {ACM} Int'l Conf. on Information and
  Knowledge Management (CIKM)}}.
\newblock


\bibitem[\protect\citeauthoryear{Chae, Kim, Chau, and Kim}{Chae
  et~al\mbox{.}}{2020}]%
        {ChaeKCK20}
\bibfield{author}{\bibinfo{person}{Dong{-}Kyu Chae}, \bibinfo{person}{Jihoo
  Kim}, \bibinfo{person}{Duen~Horng Chau}, {and} \bibinfo{person}{Sang{-}Wook
  Kim}.} \bibinfo{year}{2020}\natexlab{}.
\newblock \showarticletitle{{AR-CF:} Augmenting Virtual Users and Items in
  Collaborative Filtering for Addressing Cold-Start Problems}. In
  \bibinfo{booktitle}{\emph{Proc. of the {ACM} Int'l Conf. on Research \&
  Development in Information Retrieval (SIGIR)}}.
\newblock


\bibitem[\protect\citeauthoryear{Chen, Lin, Li, Li, Zhou, and Sun}{Chen
  et~al\mbox{.}}{2020a}]%
        {chen2020measuring}
\bibfield{author}{\bibinfo{person}{Deli Chen}, \bibinfo{person}{Yankai Lin},
  \bibinfo{person}{Wei Li}, \bibinfo{person}{Peng Li}, \bibinfo{person}{Jie
  Zhou}, {and} \bibinfo{person}{Xu Sun}.} \bibinfo{year}{2020}\natexlab{a}.
\newblock \showarticletitle{Measuring and relieving the over-smoothing problem
  for graph neural networks from the topological view}. In
  \bibinfo{booktitle}{\emph{Proc. of the AAAI Conf. on Artificial Intelligence
  (AAAI)}}.
\newblock


\bibitem[\protect\citeauthoryear{Chen, Wu, Hong, Zhang, and Wang}{Chen
  et~al\mbox{.}}{2020b}]%
        {Chen20LRGCCF}
\bibfield{author}{\bibinfo{person}{Lei Chen}, \bibinfo{person}{Le Wu},
  \bibinfo{person}{Richang Hong}, \bibinfo{person}{Kun Zhang}, {and}
  \bibinfo{person}{Meng Wang}.} \bibinfo{year}{2020}\natexlab{b}.
\newblock \showarticletitle{Revisiting Graph Based Collaborative Filtering: A
  Linear Residual Graph Convolutional Network Approach}. In
  \bibinfo{booktitle}{\emph{Proc. of the AAAI Conf. on Artificial Intelligence
  (AAAI)}}.
\newblock


\bibitem[\protect\citeauthoryear{Choi, Jeon, and Park}{Choi
  et~al\mbox{.}}{2021}]%
        {choi2021ltocf}
\bibfield{author}{\bibinfo{person}{Jeongwhan Choi}, \bibinfo{person}{Jinsung
  Jeon}, {and} \bibinfo{person}{Noseong Park}.}
  \bibinfo{year}{2021}\natexlab{}.
\newblock \showarticletitle{LT-OCF: Learnable-Time ODE-based Collaborative
  Filtering}. In \bibinfo{booktitle}{\emph{Proc. of the {ACM} Int'l Conf. on
  Information and Knowledge Management (CIKM)}}.
\newblock


\bibitem[\protect\citeauthoryear{Ebesu, Shen, and Fang}{Ebesu
  et~al\mbox{.}}{2018}]%
        {Ebesu2018}
\bibfield{author}{\bibinfo{person}{Travis Ebesu}, \bibinfo{person}{Bin Shen},
  {and} \bibinfo{person}{Yi Fang}.} \bibinfo{year}{2018}\natexlab{}.
\newblock \showarticletitle{{Collaborative Memory Network for Recommendation
  Systems}}. In \bibinfo{booktitle}{\emph{Proc. of the ACM Int'l Conf. on
  Research \& Development in Information Retrieval (SIGIR)}}.
\newblock


\bibitem[\protect\citeauthoryear{Freeman}{Freeman}{1978}]%
        {freeman1978centralitycls}
\bibfield{author}{\bibinfo{person}{Linton~C Freeman}.}
  \bibinfo{year}{1978}\natexlab{}.
\newblock \showarticletitle{Centrality in social networks conceptual
  clarification}.
\newblock \bibinfo{journal}{\emph{Social networks}} \bibinfo{volume}{1},
  \bibinfo{number}{3} (\bibinfo{year}{1978}), \bibinfo{pages}{215--239}.
\newblock


\bibitem[\protect\citeauthoryear{Goodfellow, Pouget-Abadie, Mirza, Xu,
  Warde-Farley, Ozair, Courville, and Bengio}{Goodfellow et~al\mbox{.}}{2014}]%
        {goodfellow2014generative}
\bibfield{author}{\bibinfo{person}{Ian Goodfellow}, \bibinfo{person}{Jean
  Pouget-Abadie}, \bibinfo{person}{Mehdi Mirza}, \bibinfo{person}{Bing Xu},
  \bibinfo{person}{David Warde-Farley}, \bibinfo{person}{Sherjil Ozair},
  \bibinfo{person}{Aaron Courville}, {and} \bibinfo{person}{Yoshua Bengio}.}
  \bibinfo{year}{2014}\natexlab{}.
\newblock \showarticletitle{Generative adversarial nets}. In
  \bibinfo{booktitle}{\emph{Proc. of the Annual Conf. on Neural Information
  Processing Systems (NeurIPS)}}.
\newblock


\bibitem[\protect\citeauthoryear{Gumbel}{Gumbel}{1954}]%
        {Gumbel54}
\bibfield{author}{\bibinfo{person}{Emil~Julius Gumbel}.}
  \bibinfo{year}{1954}\natexlab{}.
\newblock \showarticletitle{Statistical theory of extreme values and some
  practical applications}.
\newblock \bibinfo{journal}{\emph{NBS Applied Mathematics Series}}
  \bibinfo{volume}{33} (\bibinfo{year}{1954}).
\newblock


\bibitem[\protect\citeauthoryear{Hamilton, Ying, and Leskovec}{Hamilton
  et~al\mbox{.}}{2017}]%
        {Hamilton2017}
\bibfield{author}{\bibinfo{person}{William~L. Hamilton}, \bibinfo{person}{Rex
  Ying}, {and} \bibinfo{person}{Jure Leskovec}.}
  \bibinfo{year}{2017}\natexlab{}.
\newblock \showarticletitle{{Inductive Representation Learning on Large
  Graphs}}. In \bibinfo{booktitle}{\emph{Proc. of the Annual Conf. on Neural
  Information Processing Systems (NeurIPS)}}.
\newblock


\bibitem[\protect\citeauthoryear{He, Ye, Shen, and Zhang}{He
  et~al\mbox{.}}{2020b}]%
        {he2020milenas}
\bibfield{author}{\bibinfo{person}{Chaoyang He}, \bibinfo{person}{Haishan Ye},
  \bibinfo{person}{Li Shen}, {and} \bibinfo{person}{Tong Zhang}.}
  \bibinfo{year}{2020}\natexlab{b}.
\newblock \showarticletitle{Milenas: Efficient neural architecture search via
  mixed-level reformulation}. In \bibinfo{booktitle}{\emph{Proc. of the
  IEEE/CVF Conf. on Computer Vision and Pattern Recognition (CVPR)}}.
\newblock


\bibitem[\protect\citeauthoryear{He and McAuley}{He and McAuley}{2016}]%
        {he2016ups}
\bibfield{author}{\bibinfo{person}{Ruining He} {and} \bibinfo{person}{Julian
  McAuley}.} \bibinfo{year}{2016}\natexlab{}.
\newblock \showarticletitle{Ups and downs: Modeling the visual evolution of
  fashion trends with one-class collaborative filtering}. In
  \bibinfo{booktitle}{\emph{Proc. of The Web Conference (WWW)}}.
\newblock


\bibitem[\protect\citeauthoryear{He, Deng, Wang, Li, Zhang, and Wang}{He
  et~al\mbox{.}}{2020a}]%
        {He20LightGCN}
\bibfield{author}{\bibinfo{person}{Xiangnan He}, \bibinfo{person}{Kuan Deng},
  \bibinfo{person}{Xiang Wang}, \bibinfo{person}{Yan Li},
  \bibinfo{person}{YongDong Zhang}, {and} \bibinfo{person}{Meng Wang}.}
  \bibinfo{year}{2020}\natexlab{a}.
\newblock \showarticletitle{LightGCN: Simplifying and Powering Graph
  Convolution Network for Recommendation}. In \bibinfo{booktitle}{\emph{Proc.
  of the ACM Int'l Conf. on Research \& Development in Information Retrieval
  (SIGIR)}}.
\newblock


\bibitem[\protect\citeauthoryear{He, Liao, Zhang, Nie, Hu, and Chua}{He
  et~al\mbox{.}}{2017}]%
        {He17NeuMF}
\bibfield{author}{\bibinfo{person}{Xiangnan He}, \bibinfo{person}{Lizi Liao},
  \bibinfo{person}{Hanwang Zhang}, \bibinfo{person}{Liqiang Nie},
  \bibinfo{person}{Xia Hu}, {and} \bibinfo{person}{Tat-seng Chua}.}
  \bibinfo{year}{2017}\natexlab{}.
\newblock \showarticletitle{{Neural Collaborative Filtering}}. In
  \bibinfo{booktitle}{\emph{Proc. of The Web Conference (WWW)}}.
\newblock


\bibitem[\protect\citeauthoryear{Hu, Xie, Zheng, Liu, Shi, Liu, and Lin}{Hu
  et~al\mbox{.}}{2020}]%
        {hu2020dsnas}
\bibfield{author}{\bibinfo{person}{Shoukang Hu}, \bibinfo{person}{Sirui Xie},
  \bibinfo{person}{Hehui Zheng}, \bibinfo{person}{Chunxiao Liu},
  \bibinfo{person}{Jianping Shi}, \bibinfo{person}{Xunying Liu}, {and}
  \bibinfo{person}{Dahua Lin}.} \bibinfo{year}{2020}\natexlab{}.
\newblock \showarticletitle{Dsnas: Direct neural architecture search without
  parameter retraining}. In \bibinfo{booktitle}{\emph{Proc. of the IEEE/CVF
  Conf. on Computer Vision and Pattern Recognition (CVPR)}}.
\newblock


\bibitem[\protect\citeauthoryear{Hu, Koren, and Volinsky}{Hu
  et~al\mbox{.}}{2008}]%
        {Hu08WRMF}
\bibfield{author}{\bibinfo{person}{Yifan Hu}, \bibinfo{person}{Yehuda Koren},
  {and} \bibinfo{person}{Chris Volinsky}.} \bibinfo{year}{2008}\natexlab{}.
\newblock \showarticletitle{Collaborative filtering for implicit feedback
  datasets}. In \bibinfo{booktitle}{\emph{Proc. of the IEEE Int'l Conf. on Data
  Mining (ICDM)}}.
\newblock


\bibitem[\protect\citeauthoryear{Hwang, Parc, Kim, Lee, and Lee}{Hwang
  et~al\mbox{.}}{2016}]%
        {HwangPKLL16}
\bibfield{author}{\bibinfo{person}{Won{-}Seok Hwang}, \bibinfo{person}{Juan
  Parc}, \bibinfo{person}{Sang{-}Wook Kim}, \bibinfo{person}{Jongwuk Lee},
  {and} \bibinfo{person}{Dongwon Lee}.} \bibinfo{year}{2016}\natexlab{}.
\newblock \showarticletitle{"Told you i didn't like it": Exploiting
  uninteresting items for effective collaborative filtering}. In
  \bibinfo{booktitle}{\emph{Proc. of the {IEEE} Int'l Conf. on Data Engineering
  (ICDE)}}.
\newblock


\bibitem[\protect\citeauthoryear{Jang, Gu, and Poole}{Jang
  et~al\mbox{.}}{2016}]%
        {jang16categorical}
\bibfield{author}{\bibinfo{person}{Eric Jang}, \bibinfo{person}{Shixiang Gu},
  {and} \bibinfo{person}{Ben Poole}.} \bibinfo{year}{2016}\natexlab{}.
\newblock \showarticletitle{Categorical reparameterization with
  gumbel-softmax}. In \bibinfo{booktitle}{\emph{Proc. of the Int'l Conf. on
  Learning Representations (ICLR)}}.
\newblock


\bibitem[\protect\citeauthoryear{Kingma and Welling}{Kingma and
  Welling}{2013}]%
        {kingma2013auto}
\bibfield{author}{\bibinfo{person}{Diederik~P Kingma} {and}
  \bibinfo{person}{Max Welling}.} \bibinfo{year}{2013}\natexlab{}.
\newblock \showarticletitle{Auto-encoding variational bayes}.
\newblock \bibinfo{journal}{\emph{arXiv preprint arXiv:1312.6114}}
  (\bibinfo{year}{2013}).
\newblock


\bibitem[\protect\citeauthoryear{Kipf and Welling}{Kipf and Welling}{2017}]%
        {kipf2017semi}
\bibfield{author}{\bibinfo{person}{Thomas~N. Kipf} {and} \bibinfo{person}{Max
  Welling}.} \bibinfo{year}{2017}\natexlab{}.
\newblock \showarticletitle{Semi-Supervised Classification with Graph
  Convolutional Networks}. In \bibinfo{booktitle}{\emph{Proc. of the Int'l
  Conf. on Learning Representations (ICLR)}}.
\newblock


\bibitem[\protect\citeauthoryear{{Koren}, {Bell}, and {Volinsky}}{{Koren}
  et~al\mbox{.}}{2009}]%
        {Koren09MF}
\bibfield{author}{\bibinfo{person}{Y. {Koren}}, \bibinfo{person}{R. {Bell}},
  {and} \bibinfo{person}{C. {Volinsky}}.} \bibinfo{year}{2009}\natexlab{}.
\newblock \showarticletitle{Matrix Factorization Techniques for Recommender
  Systems}.
\newblock \bibinfo{journal}{\emph{Computer}} \bibinfo{volume}{42},
  \bibinfo{number}{8} (\bibinfo{year}{2009}), \bibinfo{pages}{30--37}.
\newblock


\bibitem[\protect\citeauthoryear{Lee, Kim, and Lee}{Lee et~al\mbox{.}}{2018}]%
        {LeeK018}
\bibfield{author}{\bibinfo{person}{Yeon{-}Chang Lee},
  \bibinfo{person}{Sang{-}Wook Kim}, {and} \bibinfo{person}{Dongwon Lee}.}
  \bibinfo{year}{2018}\natexlab{}.
\newblock \showarticletitle{gOCCF: Graph-Theoretic One-Class Collaborative
  Filtering Based on Uninteresting Items}. In \bibinfo{booktitle}{\emph{Proc.
  of the {AAAI} Conf. on Artificial Intelligence (AAAI)}}.
\newblock


\bibitem[\protect\citeauthoryear{Li, Dong, Wang, and Xu}{Li
  et~al\mbox{.}}{2020}]%
        {li2020neural}
\bibfield{author}{\bibinfo{person}{Yanxi Li}, \bibinfo{person}{Minjing Dong},
  \bibinfo{person}{Yunhe Wang}, {and} \bibinfo{person}{Chang Xu}.}
  \bibinfo{year}{2020}\natexlab{}.
\newblock \showarticletitle{Neural architecture search in a proxy validation
  loss landscape}. In \bibinfo{booktitle}{\emph{Proc. of the Int'l Conf. on
  Machine Learning (ICML)}}.
\newblock


\bibitem[\protect\citeauthoryear{Liang, Charlin, McInerney, and Blei}{Liang
  et~al\mbox{.}}{2016}]%
        {liang2016modeling}
\bibfield{author}{\bibinfo{person}{Dawen Liang}, \bibinfo{person}{Laurent
  Charlin}, \bibinfo{person}{James McInerney}, {and} \bibinfo{person}{David~M
  Blei}.} \bibinfo{year}{2016}\natexlab{}.
\newblock \showarticletitle{Modeling user exposure in recommendation}. In
  \bibinfo{booktitle}{\emph{Proc. of The Web Conference (WWW)}}.
\newblock


\bibitem[\protect\citeauthoryear{Maddison, Tarlow, and Minka}{Maddison
  et~al\mbox{.}}{2014}]%
        {Maddison14sampling}
\bibfield{author}{\bibinfo{person}{Chris~J Maddison}, \bibinfo{person}{Daniel
  Tarlow}, {and} \bibinfo{person}{Tom Minka}.} \bibinfo{year}{2014}\natexlab{}.
\newblock \showarticletitle{A* sampling}.
\newblock \bibinfo{journal}{\emph{arXiv preprint arXiv:1411.0030}}
  (\bibinfo{year}{2014}).
\newblock


\bibitem[\protect\citeauthoryear{Page, Brin, Motwani, and Winograd}{Page
  et~al\mbox{.}}{1999}]%
        {page1999pagerank}
\bibfield{author}{\bibinfo{person}{Lawrence Page}, \bibinfo{person}{Sergey
  Brin}, \bibinfo{person}{Rajeev Motwani}, {and} \bibinfo{person}{Terry
  Winograd}.} \bibinfo{year}{1999}\natexlab{}.
\newblock \bibinfo{booktitle}{\emph{The PageRank citation ranking: Bringing
  order to the web.}}
\newblock \bibinfo{type}{{T}echnical {R}eport}. \bibinfo{institution}{Stanford
  InfoLab}.
\newblock


\bibitem[\protect\citeauthoryear{Park, Yoo, Bahng, Choo, and Park}{Park
  et~al\mbox{.}}{2018}]%
        {Park2018}
\bibfield{author}{\bibinfo{person}{David~Keetae Park},
  \bibinfo{person}{Seungjoo Yoo}, \bibinfo{person}{Hyojin Bahng},
  \bibinfo{person}{Jaegul Choo}, {and} \bibinfo{person}{Noseong Park}.}
  \bibinfo{year}{2018}\natexlab{}.
\newblock \showarticletitle{{MEGAN: Mixture of Experts of Generative
  Adversarial Networks for Multimodal Image Generation}}. In
  \bibinfo{booktitle}{\emph{Proc. of the Int'l Joint Conf. on Artificial
  Intelligence (IJCAI)}}.
\newblock


\bibitem[\protect\citeauthoryear{Rendle, Freudenthaler, Gantner, and
  Schmidt-Thieme}{Rendle et~al\mbox{.}}{2009}]%
        {Rendle09BPR}
\bibfield{author}{\bibinfo{person}{Steffen Rendle}, \bibinfo{person}{Christoph
  Freudenthaler}, \bibinfo{person}{Zeno Gantner}, {and} \bibinfo{person}{Lars
  Schmidt-Thieme}.} \bibinfo{year}{2009}\natexlab{}.
\newblock \showarticletitle{BPR: Bayesian Personalized Ranking from Implicit
  Feedback}. In \bibinfo{booktitle}{\emph{Proc. of the Int’l Conf. on
  Uncertainty in Artificial Intelligence (UAI)}}.
\newblock


\bibitem[\protect\citeauthoryear{Tang, Qu, Wang, Zhang, Yan, and Mei}{Tang
  et~al\mbox{.}}{2015}]%
        {tang15line}
\bibfield{author}{\bibinfo{person}{Jian Tang}, \bibinfo{person}{Meng Qu},
  \bibinfo{person}{Mingzhe Wang}, \bibinfo{person}{Ming Zhang},
  \bibinfo{person}{Jun Yan}, {and} \bibinfo{person}{Qiaozhu Mei}.}
  \bibinfo{year}{2015}\natexlab{}.
\newblock \showarticletitle{Line: Large-scale information network embedding}.
  In \bibinfo{booktitle}{\emph{Proc. of The Web Conference (WWW)}}.
\newblock


\bibitem[\protect\citeauthoryear{van~den Berg, Kipf, and Welling}{van~den Berg
  et~al\mbox{.}}{2018}]%
        {Rianne17GCMC}
\bibfield{author}{\bibinfo{person}{Rianne van~den Berg},
  \bibinfo{person}{Thomas~N. Kipf}, {and} \bibinfo{person}{Max Welling}.}
  \bibinfo{year}{2018}\natexlab{}.
\newblock \showarticletitle{Graph Convolutional Matrix Completion}. In
  \bibinfo{booktitle}{\emph{Proc. of the ACM Int'l Conf. on Knowledge Discovery
  and Data Mining (KDD)}}.
\newblock


\bibitem[\protect\citeauthoryear{Vincent, Larochelle, Bengio, and
  Manzagol}{Vincent et~al\mbox{.}}{2008}]%
        {vincent2008extracting}
\bibfield{author}{\bibinfo{person}{Pascal Vincent}, \bibinfo{person}{Hugo
  Larochelle}, \bibinfo{person}{Yoshua Bengio}, {and}
  \bibinfo{person}{Pierre-Antoine Manzagol}.} \bibinfo{year}{2008}\natexlab{}.
\newblock \showarticletitle{Extracting and composing robust features with
  denoising autoencoders}. In \bibinfo{booktitle}{\emph{Proc. of the Int'l
  Conf. on Machine Learning (ICML)}}.
\newblock


\bibitem[\protect\citeauthoryear{Wang, Wang, and Yeung}{Wang
  et~al\mbox{.}}{2014}]%
        {Wang2014}
\bibfield{author}{\bibinfo{person}{Hao Wang}, \bibinfo{person}{Naiyan Wang},
  {and} \bibinfo{person}{Dit-Yan Yeung}.} \bibinfo{year}{2014}\natexlab{}.
\newblock \showarticletitle{{Collaborative Deep Learning for Recommender
  Systems}}. In \bibinfo{booktitle}{\emph{Proc. of the ACM Int'l Conf. on
  Knowledge Discovery and Data Mining (KDD)}}.
\newblock


\bibitem[\protect\citeauthoryear{Wang, He, Wang, Feng, and Chua}{Wang
  et~al\mbox{.}}{2019}]%
        {Wang19NGCF}
\bibfield{author}{\bibinfo{person}{Xiang Wang}, \bibinfo{person}{Xiangnan He},
  \bibinfo{person}{Meng Wang}, \bibinfo{person}{Fuli Feng}, {and}
  \bibinfo{person}{Tat-Seng Chua}.} \bibinfo{year}{2019}\natexlab{}.
\newblock \showarticletitle{Neural Graph Collaborative Filtering}. In
  \bibinfo{booktitle}{\emph{Proc. of the ACM Int'l Conf. on Research \&
  Development in Information Retrieval (SIGIR)}}.
\newblock


\bibitem[\protect\citeauthoryear{Wu, Dai, Zhang, Wang, Sun, Wu, Tian, Vajda,
  Jia, and Keutzer}{Wu et~al\mbox{.}}{2019a}]%
        {wu2019fbnet}
\bibfield{author}{\bibinfo{person}{Bichen Wu}, \bibinfo{person}{Xiaoliang Dai},
  \bibinfo{person}{Peizhao Zhang}, \bibinfo{person}{Yanghan Wang},
  \bibinfo{person}{Fei Sun}, \bibinfo{person}{Yiming Wu},
  \bibinfo{person}{Yuandong Tian}, \bibinfo{person}{Peter Vajda},
  \bibinfo{person}{Yangqing Jia}, {and} \bibinfo{person}{Kurt Keutzer}.}
  \bibinfo{year}{2019}\natexlab{a}.
\newblock \showarticletitle{Fbnet: Hardware-aware efficient convnet design via
  differentiable neural architecture search}. In
  \bibinfo{booktitle}{\emph{Proc. of the IEEE/CVF Conf. on Computer Vision and
  Pattern Recognition (CVPR)}}.
\newblock


\bibitem[\protect\citeauthoryear{Wu, Zhang, de~Souza, Fifty, Yu, and
  Weinberger}{Wu et~al\mbox{.}}{2019b}]%
        {Wu2019SGC}
\bibfield{author}{\bibinfo{person}{Felix Wu}, \bibinfo{person}{Tianyi Zhang},
  \bibinfo{person}{Amauri~Holanda de Souza}, \bibinfo{person}{Christopher
  Fifty}, \bibinfo{person}{Tao Yu}, {and} \bibinfo{person}{Kilian~Q.
  Weinberger}.} \bibinfo{year}{2019}\natexlab{b}.
\newblock \showarticletitle{{Simplifying Graph Convolutional Networks}}. In
  \bibinfo{booktitle}{\emph{Proc. of the Int'l Conf. on Machine Learning
  (ICML)}}.
\newblock


\bibitem[\protect\citeauthoryear{Ying, He, Chen, Eksombatchai, Hamilton, and
  Leskovec}{Ying et~al\mbox{.}}{2018}]%
        {Ying18pinsage}
\bibfield{author}{\bibinfo{person}{Rex Ying}, \bibinfo{person}{Ruining He},
  \bibinfo{person}{Kaifeng Chen}, \bibinfo{person}{Pong Eksombatchai},
  \bibinfo{person}{William~L. Hamilton}, {and} \bibinfo{person}{Jure
  Leskovec}.} \bibinfo{year}{2018}\natexlab{}.
\newblock \showarticletitle{Graph Convolutional Neural Networks for Web-Scale
  Recommender Systems}. In \bibinfo{booktitle}{\emph{Proc. of the ACM Int'l
  Conf. on Knowledge Discovery and Data Mining (KDD)}}.
\newblock


\bibitem[\protect\citeauthoryear{Zhang, Liu, and Jin}{Zhang
  et~al\mbox{.}}{2020}]%
        {zhang2020survey}
\bibfield{author}{\bibinfo{person}{Guijuan Zhang}, \bibinfo{person}{Yang Liu},
  {and} \bibinfo{person}{Xiaoning Jin}.} \bibinfo{year}{2020}\natexlab{}.
\newblock \showarticletitle{A survey of autoencoder-based recommender systems}.
\newblock \bibinfo{journal}{\emph{Frontiers of Computer Science}}
  \bibinfo{volume}{14}, \bibinfo{number}{2} (\bibinfo{year}{2020}),
  \bibinfo{pages}{430--450}.
\newblock


\end{thebibliography}



\end{document}